\documentclass[aps,reprint,superscriptaddress,nofootinbib,showpacs,amsmath,amssymb]{revtex4-1}
\usepackage{latexsym,graphicx,slashed,bm,color,dcolumn}
\usepackage{hyperref}

\newcommand{\be}[1]{\begin{equation}\label{#1}}
\newcommand{\beq}{\begin{equation}}
\newcommand{\eeq}{\end{equation}}
\def\ee{\end{equation}}
\newcommand{\beqn}[1]{\begin{eqnarray}\label{#1}}
\newcommand{\eeqn}{\end{eqnarray}}

\newcommand{\mat}[4]{\left(\begin{array}{cc}{#1}&{#2}\\{#3}&{#4}
\end{array}\right)}

\renewcommand{\to}{\rightarrow}

\def\ov{\overline}

\def\lsim{\raise0.3ex\hbox{$\;<$\kern-0.75em\raise-1.1ex
\hbox{$\sim\;$}}}
\def\gsim{\raise0.3ex\hbox{$\;>$\kern-0.75em\raise-1.1ex
\hbox{$\sim\;$}}}
\def\cal{\mathcal}
\def\cF{{\cal F}}

\def\cL{{\cal L}}
\def\cM{{\cal M}}

\def\eps{\varepsilon}

\def\rB{{\rm B}}
\def\rL{{\rm L}}

\begin{document}

\title{Neutron--mirror neutron mixing and neutron stars  }

\author{Zurab~Berezhiani}
\affiliation{Dipartimento di Fisica e Chimica, Universit\`a di L'Aquila, 67100 Coppito, L'Aquila, Italy} 
\affiliation{INFN, Laboratori Nazionali del Gran Sasso, 67010 Assergi,  L'Aquila, Italy}

\author{Riccardo Biondi}
\affiliation{INFN, Laboratori Nazionali del Gran Sasso, 67010 Assergi,  L'Aquila, Italy}

\author{Massimo Mannarelli}
\affiliation{INFN, Laboratori Nazionali del Gran Sasso, 67010 Assergi,  L'Aquila, Italy}

\author{Francesco Tonelli}
\affiliation{Consiglio Nazionale delle Ricerche, CNR-SPIN, 67100 Coppito, L'Aquila, Italy}


\begin{abstract} 
The oscillation of  neutrons $n$ into mirror  neutrons $n'$,  their mass degenerate 
partners from  dark mirror sector,  can have interesting implications for neutron stars: 
an ordinary neutron star could gradually transform  into a mixed star consisting in part of mirror dark matter.  
Mixed stars can be detectable as  twin partners of ordinary neutron stars:   
namely, there can exist compact stars with the same masses but having different radii. 
For a given equation of state (identical between the ordinary and mirror components), 
the mass and radius of  a mixed star depend on the proportion between
 the ordinary and mirror components in its interior which in turn depends on its age. 
 If $50 \% - 50\%$ proportion  between two fractions can be reached asymptotically in time, 
 then  the maximum mass of such ``maximally mixed stars"  should be  $\sqrt2$ times smaller 
than that of ordinary neutron star while the stars exceeding a critical mass value $M^{\rm max}_{NS}/\sqrt2$ 
should collapse in black holes after certain time.   
We evaluate the evolution time and  discuss the implications of $n-n'$ transition 
for the pulsar observations as well as for the gravitational waves from the neutron star mergers 
and associated electromagnetic signals. 

\end{abstract}

\maketitle



 \section{Introduction}  
 
The idea that there may exist a hidden particle sector consisting of mirror duplicates of
the observed standard particles    
was introduced long time ago for restoring parity as a fundamental symmetry \cite{Mirror}.  
In this context, all known particles: electron $e$, proton $p$, neutron $n$,  
neutrinos $\nu$ etc. must have 
the mirror partners  $e'$, $p'$, $n'$, $\nu'$ etc. which 
are supposed to be sterile to our Standard Model (SM) interactions $SU(3)\times SU(2)\times U(1)$, 
but have their own $SU(3)'\times SU(2)'\times U(1)'$ gauge interactions 
(see e.g. \cite{mirror,Alice,Foot} for reviews and \cite{Okun} for a historical overview). 

More generically, one can consider theories  based on the direct product $G\times G'$ of two 
identical gauge factors (SM or some its extension),   
with  Lagrangian  
\be{Lagr}
{\cal L}_{\rm tot} = {\cal L} + {\cal L}' + {\cal L}_{\rm mix}  
\ee
 where ${\cal L}$  describes ordinary (O) particles, ${\cal L}'$ describes their mirror (M) partners 
 and ${\cal L}_{\rm mix}$ stands for possible cross-interactions between the O and M particles.  
 The identical forms of the Lagrangians ${\cal L}$ and ${\cal L}'$ is ensured by  the
$Z_2$ symmetry under the exchange $G\leftrightarrow G'$ of all O particles 
(fermions, Higgs and gauge fields) with their M twins (`primed' fermions, Higgs and gauge fields). 
In view of chiral character of the SM, this discrete symmetry can be imposed {\it with} or {\it without} 
chirality change between O and M fermions. Mirror parity corresponds to the former possibility,    
however this difference will not be important for our further considerations. 

M matter, invisible in terms of our photons, is gravitationally coupled  to O matter and
can contribute to cosmological dark matter, or perhaps could even  represent its entire amount. 
It can form  mirror nuclei and atoms 
with an abundance related to the mirror baryon asymmetry.  
The  collisional  and dissipative properties of M baryons can have   
specific implications for cosmology 
\cite{Khlopov,BDM,BCV,BCCV,BCCP} as well as for the dark matter direct search \cite{Cerulli}. 

If  $Z_2$ symmetry is unbroken,  then O particles  and their M partners are degenerate 
in mass, and the interaction terms in ${\cal L}$ and  ${\cal L}'$ are exactly identical.  
Also the case of spontaneously broken  $Z_2$ symmetry has been considered in the literature~\cite{BDM}
when the weak scales in the two sectors become different, 
and thus mass splittings emerge between M particles and their O partners. 

Lagrangian  ${\cal L}_{\rm mix}$ in \eqref{Lagr} 
may contain cross-terms that induce  oscillation phenomena between O and M sectors. 
In fact, any neutral O particle, elementary (e.g. neutrinos) or composite (e.g.  neutrons)  
can have mixing with its M twin and oscillate into the latter.   
In particular, ``active-sterile" neutrino mixing between ordinary $\nu_{e,\mu,\tau}$ 
and mirror $\nu'_{e,\mu,\tau}$ can be induced 
via effective operators $\frac{1}{M} \ell \ell' \phi \phi' +{\rm h.c.}$ 
with a large cutoff scale  $M$, 
where $\ell$ and $\phi$ are lepton and Higgs doublets of O sector and 
$\ell'$ and $\phi'$ are their mirror partners   \cite{ABS,FV}. 
These operators, violating both $\rB-\rL$ and $\rB'-\rL'$ symmetries,  
also suggest a co-leptogenesis mechanism  which creates 
comparable baryon asymmetries in both O and M worlds, 
with $\Omega'_{\rm B}\geq \Omega_{\rm B}$  \cite{BB-PRL,EPJ-ST,ADM-IJMP}.  


In this paper we shall concentrate on the mixing between the neutron 
$n$ and its mirror partner $n'$ \cite{BB-nn',More}: 
\be{n-npr}
\eps \, \ov{n'} n + \text{h.c.}
\ee
In fact, the mixed Lagrangian $\cL_\text{mix}$ may include 
effective operators  involving color-singlet 
combinations of ordinary $u,d$ and mirror  $u',d'$ quarks:
 \be{operator}
\frac{1}{\cM^5}(\ov{u'}\,\ov{d'}\,\ov{d'})(udd)  + {\rm h.c.}\,, 
\ee
(the gauge and Lorentz indices are omitted). 
These operators violate both O and M baryon numbers by one unit,  $\Delta\rB=1$ and $\Delta\rB'=-1$, 
but the overall baryon number $\ov{\rB}=\rB+\rB'$ is conserved.
In UV complete theories they can be induced via a see-saw like mechanism involving 
new heavy particles, colored scalars and neutral fermions, with masses $\sim \cM$ \cite{BB-nn',B-M}. 
Hence, for $\cM$ at few TeV, the underlying theories can be testable 
at the LHC and future accelerators~\cite{B-M,Fajfer}.  

Operators \eqref{operator} induce $n-n'$ mass mixing \eqref{n-npr} 
with 
\be{eps} 
\eps = \frac{C^2 \Lambda_{\rm QCD}^6}{\cM^5} = C^2 \left(\frac{10~{\rm TeV}}{\cM}\right)^5 
\times 10^{-15}~{\rm eV} 
\ee
where $C=O(1)$ is the operator dependent numerical factor 
 in the determination of the matrix element  $\langle 0\vert udd \vert n \rangle$.  

The phenomenon of $n-n'$ oscillation is analogous to that of neutron-antineutron ($n-\bar n$) 
oscillation \cite{Kuzmin} 
(for a review, see \cite{Phillips}), and  in fact both phenomena 
can be related to the same new physics \cite{B-M,B-L}. 
However, $n-\bar n$ oscillation is strongly restricted by experiment. 
Namely, the  direct experimental limit on $n-\bar n$ oscillation is 
$\eps_{n\bar n} < 7.7 \times 10^{-24}$~eV while the nuclear stability bounds 
are yet stronger  yielding  $\eps_{n\bar n} < 2.5 \times 10^{-24}$~eV \cite{Phillips}. 
As for $n-n'$ oscillation, it is kinematically forbidden for neutrons bound in nuclei, 
simply by energy conservation \cite{BB-nn'}, and so nuclear stability gives no 
limit on $n-n'$ mixing. 


For free neutrons $n-n'$ oscillation can be effective, and 
it may even be much faster than the neutron decay \cite{BB-nn',More}. 
Namely, $n-n'$  mixing mass can be as large as $\eps \sim 10^{-15}$~eV, 
corresponding to the characteristic oscillation time $\tau_{nn'}=\eps^{-1}$  
as small as 1~second or even smaller.  
This possibility is not excluded by existing astrophysical and cosmological limits \cite{BB-nn'},   
predicting observable effects  for the ultra-high energy cosmic rays  \cite{UHECR}, 
for neutrons from solar flares \cite{Nasri} and for primordial nucleosynthesis \cite{Coc}.   
Its search via the neutron disappearance ($n\to n'$) and regeneration 
($n \to n' \to n$) experiments can be perfectly accessible 
at existing neutron  facilities  \cite{BB-nn',Pokot,HFIR}. 

The reason why the disappearance of free neutrons 
so far skipped the experimental detection might be that in normal experimental conditions    
the $n-n'$ oscillation probability $P_{nn'}$ is suppressed by environmental effects  
as e.g.  mirror magnetic fields at the Earth \cite{More}.  
Several dedicated experiments were performed searching $n-n'$ oscillations 
\cite{Ban,Serebrov1,Altarev,Bodek,Serebrov2,ILL,Abel} and they still do not exclude  
small oscillation times if these environmental effects are strong enough.  
Moreover,  some of these experiments show anomalous deviations 
from null hypothesis indicating to $\tau_{nn'} \sim 10$~s or so \cite{ILL,Nesti}. 
New experiments for testing  these effects are underway~\cite{Broussard,ESS}. 

Larger values of $\eps$ are also allowed if $n$ and $n'$ are not exactly mass-degenerate. 
Moreover, $n-n'$ oscillation with $\eps \sim 10^{-10}$~eV  or so 
can solve the neutron lifetime problem, 
 the $4\sigma$ discrepancy between the neutron lifetimes 
measured via the bottle and beam experiments,  provided that
$n$ and $n'$ have a mass splitting $m_{n'}-m_n\sim 100$~neV \cite{lifetime}.  
Such a small splitting can be naturally 
realized in models in which $Z_2$ parity is spontaneously broken \cite{BDM} but with a
 rather small difference between the O and M Higgs VEVs $\langle \phi \rangle$ and $\langle \phi' \rangle$
\cite{Nussinov}.  

Although $n-n'$ transition is forbidden for neutrons bound in nuclei by nuclear forces, it  
is allowed in neutron stars (NS) in which neutrons are bound gravitationally. 
The transformation of nuclear matter into mirror matter should decrease the  degeneracy pressure, 
thereby softening the equation of state (EoS) of the system. 
 The gravitational binding energy increases and thus the process  is energetically favorable. 
By $n-n'$ transitions, a  NS  born after supernova explosions gradually evolves in a mixed star (MS)  
partially  consisting of mirror matter,  with decreasing  gravitational mass and radius. 
The softening of the EoS is particularly effective for the NS born with large masses. 
In this case, the conversion of O matter in M matter may produce a  gravitationally 
unstable MS that collapses to a black hole (BH). On the other hand, a NS born with a mass below 
a threshold value would asymptotically evolve in maximally mixed star (MMS),  
with equal amounts of the O and M components.  

In the present paper we shall concentrate on the implications of $n-n'$ conversion for the NS.
As was already noted in~\cite{BB-nn'}, for $\eps \lesssim 10^{-15}$~eV (i.e. $\tau_{nn'} \sim 1$~s or so) 
the transformation time  is several orders of magnitude larger than the age of the Universe.  
Nevertheless, we show that the NS transformation into the MS with small mirror cores can have 
astrophysical signatures for the pulsar dynamics and for the gravitational mergers. 
On the other hand, as discussed above,  larger values of $\eps$ are not excluded and the   
conversion process can be  more rapid,
 so that the older NS could be already transformed in the MMS.
The respective implications for  NSs 
were briefly discussed in \cite{INT}, in more details in \cite{Massimo},  
and were recently addressed also in \cite{Nussinov-new,Ciancarella}.
Here we analyse conditions for the MS stability, derive the mass--radii scaling relations 
and discuss the possible observational effects.

Our work is organized as follows. 
In Sect.~\ref{sec:M-R} we discuss effects of $n-n'$ mixing  
on the evolution as well as on the gravitational stability of compact stars,  
and derive the mass-radius scaling relations between the  NS and MMS.  
In Sect.~\ref{sec:evolution} we discuss $n-n'$ oscillations  in dense nuclear matter 
and estimate the transformation time of neutron stars into mixed stars. 
 In Sect. \ref{sec:pulsars} we discuss possible effects for the pulsar timing observations,  
 and in Sect. \ref{sec:GW} for the neutron star coalescences and associated signals. 
 We draw our conclusions in Sec.~\ref{sec:conclusions}.


\section{Neutron star evolution in mixed stars}
\label{sec:M-R}

Neutron stars  are presumably born after the supernova explosions of massive stars.  
They  are believed to have an onion-like structure that schematically 
consists of  a thin rigid crust at the surface and a liquid nuclear matter 
in the core whose dominant component are  neutrons, 
plus  some fractions of  protons and electrons 
and perhaps also heavier baryons and muons (for reviews, see e.g. \cite{Lattimer,Vidana}). 
 
The  supernova core-collapse should produce a NS  consisting  of ordinary nuclear matter.  
However, if $n\to n'$ transitions are allowed they start to produce  M neutrons, 
and the original NS will be gradually transformed into a MS
 with  increasing fraction of mirror nuclear matter 
in its interior. 
We consider that this process conserves the total number of baryons, and 
assume that it is rather slow,  
with the effective $n-n'$ conversion rate  much less than the typical  cooling rate.   
Under these circumstances the transformation process is adiabatic and 
can be described  by the Boltzmann equations 
\be{Boltzmann}
\frac{dN_\text{O}}{dt} = - \Gamma N_\text{O} + \Gamma' N_\text{M} \, , 
\quad\quad   \frac{dN_\text{M} }{dt} = \Gamma N_\text{O} - \Gamma' N_\text{M}\,, 
\ee
where $N_\text{O}(t)$ and $N_\text{M}(t)$ are respectively the total numbers of O and 
M baryons in the star at the time $t$ and $\Gamma(t)$ is the $n\to n'$ conversion rate    
(to be estimated in next section), whereas $\Gamma' \ll \Gamma$ 
is the rate of the inverse process $n' \to n$.  
%
Starting at $t=0$ from a newborn NS  with $N_\text{O}=N$ and  $N_\text{M}=0$,   
then $N_\text{M}(t)$ will increase and $N_\text{O}(t)$ will decrease in time,  
but with  $N_\text{O}(t)+N_\text{M}(t)=N$ at any stage 
since the overall number of O and M baryons in the MS is conserved.\footnote{This is because 
$n-n'$ oscillations conserve the combined baryon number  $\ov{\rB}=\rB+\rB'$. 
In principle, the neutron could have mixings with both M neutron $n'$ and M antineutron $\ov{n}'$.  
In this case $\ov{B}$ would be violated \cite{shortcut}, but here we do not 
discuss this possibility.} 
Thus, neglecting the inverse reaction rate $\Gamma'$,  Eqs.~\eqref{Boltzmann} 
reduce to the single equation
\be{Boltzmann-X}
\frac{dX }{dt} = \Gamma \, (1- X)\,,   
\ee
where $X(t)=N_\text{M}(t)/N$ is the fraction of M baryons at the time $t$ while the 
fraction of O baryons is $N_\text{O}(t)/N=1-X(t)$. 
Asymptotically in time the star can evolve (provided that it remains stable during the evolution) 
 to the final equilibrium configuration with 
$X = 1/2$, corresponding to  the MMS     
with equal amounts of the O and M components in its interior. 

\subsection{Structure of mixed neutron stars}
\label{sec:structure}

As far as the evolution is adiabatic, we can use a static ``two fluid" description 
in which the total energy density and pressure can be decomposed as the sum 
of ordinary and mirror components, i.e. $\rho=\rho_\text{O}+\rho_\text{M}$ and $p=p_\text{O}+p_\text{M}$. 
Thus, the  total energy-momentum tensor $T^\mu_{\phantom{\mu}\nu}= {\rm diag}(\rho, -p, -p , -p )$ 
can be split as $T_{\mu\nu}=T_{\mu\nu}^{\text{O}}+T_{\mu\nu}^{\text{M}}$, where $T_{\mu\nu}^{\text{O}}$ and  $T_{\mu\nu}^{\text{M}}$ respectively are the energy momentum tensors of O and M matter. 
Assuming  spherical symmetry, at any moment of time the MS can be described 
as a static  configuration of two concentric O and M  spheres  
(we neglect the star rotation which is known to have a small effect on the mass/radius relation) 
having the radii $R_\text{O}$ and $R_\text{M}$, respectively corresponding to the   
positions where   $p_\text{O}(r)$ and $p_\text{M}(r)$  vanish. 
Hence, we take the metric  tensor   
in standard  form with spherical symmetry, 
$g_{\mu\nu}=\text{diag}(- g_{tt}, g_{rr}, r^2, r^2 \sin^2\theta)$, where \footnote{In this section 
we use geometrized units, $c=1$ and $G=1$.}
\be{g-metric}
g_{tt}(r) =  \exp[-2\phi(r)], \qquad g_{rr}(r) = \frac{1}{1-2m(r)/r } 
\ee
 where $m(r)$ is the total gravitational mass within the radius $r$,   and $\phi(r)$ is  the gravitational potential.
In hydrostatic equilibrium the density and pressure profiles in the star
are determined 
by  the Tolman-Openheimer-Volkoff (TOV) equations \cite{Tolman,OV}: 
\begin{align}
 \frac{d m}{d r} &= 4 \pi  \rho r^2   \label{eq:dm} \\ 
 \frac{d p}{d r}      & =- (\rho+p) \frac{d \phi}{d r}   \label{eq:TOV0} \\
\frac{1}{\rho+p} \frac{d p}{d r}   & = \frac{m + 4 \pi p r^3}{ 2 mr -r^2 }    \label{eq:dp} 
\end{align}
The first differential equation above is linear,  
and we can split it between two components,  $m(r)=m_\text{O}(r) + m_\text{M}(r)$: 
\be{eq:dmalpha}
\frac{d m_\text{O}}{dr} = 4 \pi  \rho_\text{O} r^2\,, \quad  \quad \frac{d m_\text{M}}{dr} = 4 \pi  \rho_\text{M} r^2 
\ee
which give   $m_\alpha(r) = 4\pi \int_0^r \rho_\alpha(r) r^2 dr$ ($\alpha= \text{O, M}$). 
Therefore, the total gravitational mass of the MS is the sum of gravitational masses 
of the two components,  $M_\text{MS}=M_\text{O}+M_\text{M}$.  
 Since both components are in hydrostatic equilibrium, 
$\dot{\rho}_\alpha,\dot{p}_\alpha=0$,  the continuity equation 
 for the energy-momentum tensor $\nabla_\mu T^{\mu(\alpha)}_\nu =0$ for each 
 of them  separately  gives
\be{eq-phi}
\partial_r\phi = \frac{\partial_r p_\text{O}} {\rho_\text{O}+p_\text{O}} 
=\frac{\partial_r p_\text{M}} {\rho_\text{M}+p_\text{M}}  = \frac{\partial_r p} {\rho+p}\,, 
 \ee
where the last equality follows from the first two.

As for Eq.~\eqref{eq:dp}, it in fact couples the pressures and energy 
densities of the two fluids.  
Using Eqs. (\ref{eq:dmalpha}) and (\ref{eq-phi}), it gives the coupled 
differential equations for the O and M components: 
\be{eq:system}
\frac{\partial_r p_\text{O}} {\rho_\text{O}\!+\!p_\text{O}} 
=\frac{\partial_r p_\text{M}} {\rho_\text{M}\!+\!p_\text{M}}  = 
 \frac{m_\text{M}\!+\!m_\text{O} +\! 4 \pi (p_\text{M}\!+\!p_\text{O}) r^3}{2 (m_\text{M}\!+\!m_\text{O}) r -r^2} 
\ee
The above system of differential equations can be solved once the EoS  
are given, and the appropriate boundary conditions are chosen. 
In our case, both  components $\alpha=\text{O,M}$ should be the same EoS 
$p_\alpha = F(\rho_\alpha)$ by mirror symmetry.  
Then one can find their density profiles $\rho_\alpha(r)$ 
by fixing the respective  central densities, 
$\rho_\text{O}(0)=\rho_\text{cO}$ and $\rho_\text{M}(0)=\rho_\text{cM}$.     
 
 For initial configuration of NS composed exclusively of O baryons, 
the system of equations~\eqref{eq:dmalpha} and \eqref{eq:system} 
reduces to the standard one-component TOV equations: 
\begin{equation}\label{TOV-NS} 
 \frac{d m_\text{O}}{d r} = 4 \pi  \rho_\text{O} r^2 , \quad 
\frac{1}{\rho_\text{O}+p_\text{O}}\,\frac{d p_\text{O}}{d r} 
= \frac{m_\text{O} + 4 \pi p_\text{O} r^3}{2 m_\text{O} r -r^2} 
\end{equation}
 Solving these equations with a given central density 
$\rho_\text{O}(0)=\rho_c$,  one can find  the NS density profile: 
\begin{equation}\label{f-NS}
\rho(r) = \rho_\text{O}(r)  = \rho_c f(r) 
\end{equation} 
where the function $f(r)$ is normalized as $f(0)=1$ and its shape   
depends on the  EoS.   
The NS radius $R_{\rm NS}\equiv R_\text{O}$ 
corresponds to the distance at which $f(r)$ vanishes,  
while the gravitational mass is
\begin{equation}\label{M-NS}
 M_{\rm NS}(\rho_\text{c})  = 4\pi \int_0^{R_{\rm NS}}\! \!\! \rho(r) r^2 dr  
 = 4\pi \rho_c \int_0^{\infty} \!\!\! f(r) r^2 dr 
\end{equation}
In fact, the integration can be extended to infinity since $f(r)=0$ for $r> R_\text{NS}$. 


An important feature is that, for any EoS, 
there is no gravitationally stable solution if $\rho_\text{c}$ exceeds 
a certain critical value $\rho_c^\text{max}$, 
which determines the maximum mass 
of the neutron stars, 
$M_{\rm NS}^{\rm max}= M_{\rm NS}(\rho_c^{\rm max})$, 
also called the {\it last stable configuration} for a given EoS. 
Discovery of pulsars  
with gravitational masses $M\approx 2 M_\odot$ \cite{Demorest,Antoniadis,Cromartie} 
challenged several EoSs excluding the too soft ones. 
Moreover, observation of PSR J1748-2021B with mass estimated 
as $M = (2.74 \pm 0.21) M_\odot$~\cite{Freire} indicates that the maximum mass 
may be considerably larger than $2M\odot$. 

In this respect,  in the following   we consider two possible EoSs 
as examples of those which allow large enough maximum masses. 
One is the realistic  SLy  (Skyrme-Lyon) EoS \cite{Sly}, which gives both 
the energy density and pressure for discrete values of the baryon number density $n_\text{O}$. 
The last stable configuration associated to this EoS has a mass 
$M^{\rm max}_{\rm NS} \simeq 2.05 \, M_{\odot}$ 
(corresponding to the baryon number $N^{\rm max}_{\rm NS} \simeq 2.91 \times 10^{57}$),
and it is reached for $\rho_c^{\rm max} \simeq 2.86 \times 10^{15}$~g/cm$^3$  
($n_\text{c}^{\rm max} \simeq 1.28$~fm$^{-3}$) \cite{Sly}. 

As a second example we choose an EoS consisting of two joined polytropes: 
\beq
p = \begin{cases}  K_1 \rho^{\gamma_1}  & \text{for}~ \rho > \rho_\text{tr}\\
 K_2 \rho^{\gamma_2}  & \text{for} ~ \rho < \rho_\text{tr}
\end{cases} 
\eeq
with $\gamma_1=3$ for the inner part of the NS and $\gamma_2=4/3$ for its outer part,  
with transition  at half the nuclear saturation density 
$\rho_\text{tr}  = 1.35 \times 10^{14}$~g/cm$^3$, for details see Ref. \cite{Lattimer}.
This EoS allows the larger maximum mass $M^{\rm max}_{\rm NS} = 2.57 \, M_{\odot}$. 
For the typical NS with $M\simeq 1.4~M_\odot$ both of these EoS imply
a radius of about $12$~km. 

Although we shall restrict to these two EoSs, there is a large number of possible EoS 
(for reviews see e.g. Refs.~\cite{Ozel,Burgio}.) 
For instance, the phenomenological EoS suggested  in Ref.~\cite{Mueller}  is stiffer,  
predicting the last stable configuration with mass as large as $2.7\, M_\odot$, and the NS 
radii can be as large as 15~km.  
 From a very fundamental point of view, without appealing to a particular  EoS   
but only requiring the micro-stability and causality conditions  
$0 < dp/d\rho < 1$, one can put an absolute upper 
limit on the NS mass, $M_{\rm NS}^{\rm max} \lesssim 3.2~M_\odot$, 
known as Rhoades-Ruffini bound~\cite{Ruffini}. 

The baryon number density $n=n_\text{O}$ is directly related to the gravitational density. 
The relation between the two densities, $n= n(\rho) $,  
depends on the chosen EoS. 
(For relevant densities the ratio $n/\rho$ is roughly constant  but not exactly.) 
Therefore, the baryon density profile can be presented  
as $n(r) = n[\rho(r)]$  
and, integrating over  the NS volume, we obtain its total baryon number:  
\be{B-NS} 
N_\text{NS}(\rho_c) = 4\pi \int_0^\infty \! g_{rr}(r)^{1/2} n\big[\rho_c f(r)\big] \, r^2 dr  
\ee
For fixed $\rho_c$, 
the shapes of $f(r)$ and $g_{rr}(r)$  depend on the EoS. 
E.g.  for the Sly EoS $\rho$ is explicitly given  as a function of $n$  in Ref.~\cite{Sly}. 
 The total baryon number of the NS depends on its mass nearly linearly: 
\be{M-R-Sly} 
N_\text{NS}= \kappa N_\odot  \, (M_\text{NS}/M_\odot)  
\ee
where $N_\odot = M_\odot/m_n = 1.188 \times 10^{57}$ is the baryon amount in the sun. 
E.g. for the Sly EoS we have $\kappa\approx 1.1$ for the typical masses $M_\text{NS} \simeq 1.4\,M_\odot$, 
which increases to $\kappa\approx 1.2$ for heavy NS with $M_\text{NS}$ approaching  $2\,M_\odot$. 
The equivalent baryonic mass is 
$M_B = m_n N_\text{NS}  =\kappa M_\text{NS}$,
so that $M_\text{NS}/M_B \approx 0.9$.   
The mass deficit $M_B - M_\text{NS} \approx 0.1 M_\text{NS}$ 
corresponds to the gravitational binding energy. 
For the stiffer  EoS the NS binding energies are smaller. 

Considering now the NS gradual conversion into a MS with $N_\text{M} < N_\text{O}$, 
we find that  $\rho_\text{O}(r) > \rho_\text{M}(r)$
at any position and at any time,  and so 
 $M_\text{O} > M_\text{M}$ and $R_\text{O} > R_\text{M}$. 
 Therefore, the radius of the mixed star is the radius of the ordinary component, 
 $R_\text{MS} = R_\text{O}$.   
 For this reason, when solving the differential equations  in Eq.~\eqref{eq:system}, 
 one has to take into account that at $r =R_\text{M}\le R_\text{O}$ 
 the  pressure of M component vanishes  
 meaning that its mass  is saturated, i.e. $m_\text{M}(r >R_\text{M}) = 
 M_\text{M}$. Thus,  at radius $R_\text{M}$ one has to stop 
 the iteration for the differential equation of M component while continuing the integration 
 of the O component equation up to the radius $R_\text{O}$.  
The maximally mixed configuration between the two components 
can be reached only asymptotically in time 
(provided that the star is enough light  to remain  stable up to the MMS configuration, see below). 
In this case the density profiles of the O and M components should be identical by symmetry, 
$\rho_\text{O}(r) = \rho_\text{M}(r)$, and thus for the MMS radius and mass we have  respectively
$R_\text{MMS}=R_\text{O}=R_\text{M}$ and $M_\text{MMS}=2M_\text{O}=2M_\text{M}$.  

It is useful to characterize the different MS configurations by  the  ratio 
of ordinary to mirror central densities, $\chi = \rho_\text{cM}/\rho_\text{cO}$. 
As far as the evolution is adiabatic, and the total baryon number is conserved,  
for a star with a total baryon number $N=N_\text{O} + N_\text{M}$ 
 the value of $\chi=\chi(X)$  is determined  by the mirror baryon fraction $X=N_\text{M}/N$ 
 at a given stage of its evolution.  
In our picture,  a newborn NS has $\chi=0$, but  $\chi$  increases with time  
when it evolves as a MS, 
and asymptotically approaches $\chi=1$ for a MMS configuration ($X=1/2$)
if the star remains stable during the evolution. 

\begin{figure}
\includegraphics[width=0.45\textwidth]{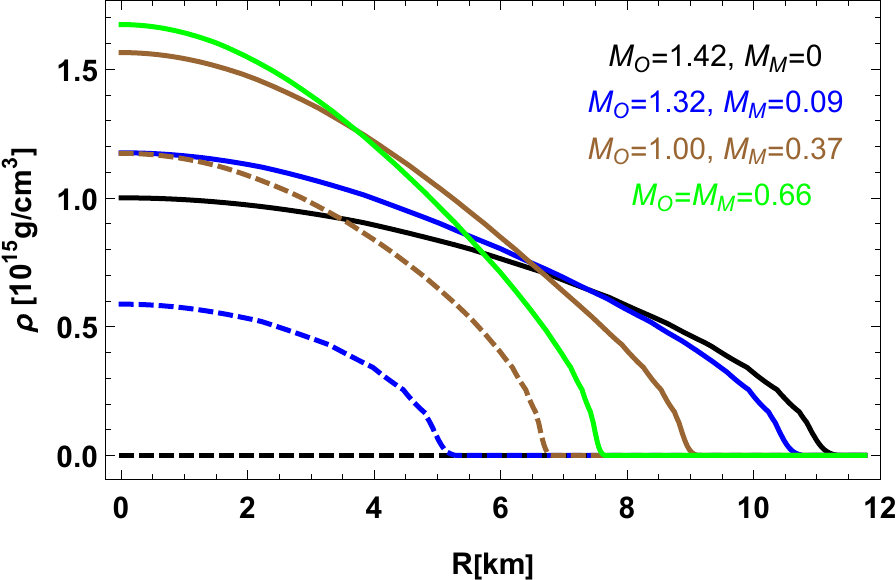}  
\caption{ \label{fig:profiles}
Density profiles $\rho_\text{O}(r)$ (solid)  and $\rho_\text{M}(r)$  (dashed) 
in a star with the initial mass $M_\text{O}\simeq1.42~M_\odot$ at different stages of its evolution 
from $\chi=0$ (black) to $\chi=1$ (green)  with 
intermediate values $\chi=0.5$ and $0.75$ (respectively,  blue and brown).
 The corresponding total masses of the ordinary and mirror spheres, respectively 
 $M_\text{O}$ and $M_\text{M}$, 
are given in the legend (in solar mass units~$M_\odot$).  
Computations are performed using the Sly EoS \cite{Sly}. 
}
\end{figure}

In Fig.~\ref{fig:profiles}, as an example, we show the O and M density profiles obtained  
using the SLy EoS at different stages of the evolution at fixed baryon number 
($N =N_\text{O} + N_\text{M}= 1.88 \times 10^{57}$), starting from the NS ($\chi=0$) 
with the initial mass $M_\text{NS}=M_\text{O}\approx 1.42~M_\odot$ 
 and radius $R_{\rm NS} = R_\text{O}\approx 12$~km. 
With increasing M fraction the gravitational mass and the radius  gradually decrease, and 
when the star becomes a MMS  ($X=1/2$, $\chi=1$) the gravitational mass  reduces to 
$M_\text{MMS}\simeq 1.32~M_\odot$ while the radius shrinks  to 
$R_\text{MMS} = R_\text{O,M} \simeq 8.5$ km. 
Hence, in the evolution the stars binding energy  has increased by about 
$0.1 M_\odot$, and its radius has decreased by about $3.5$~ km, 
with  the increase of compactness by $30 \%$ or so. 


\begin{figure}
\includegraphics[width=0.45\textwidth]{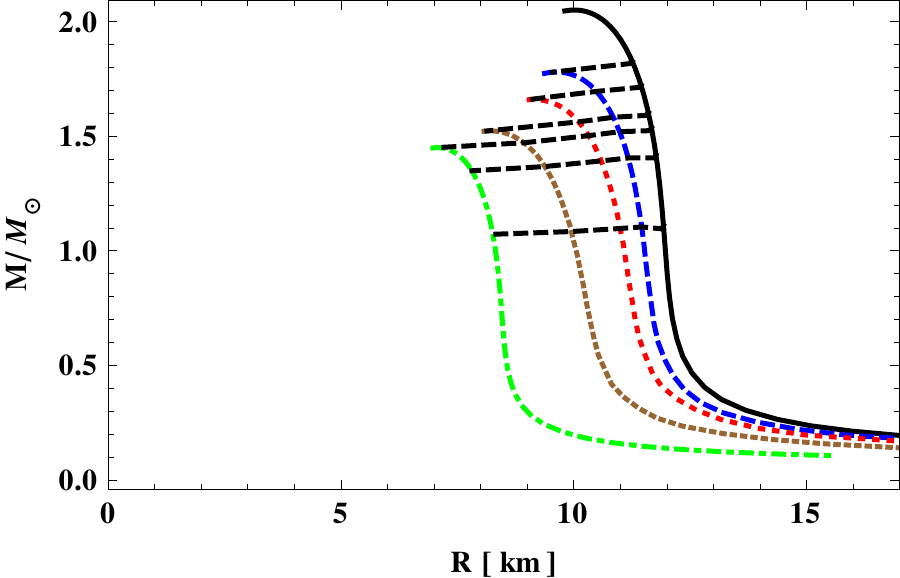} \\
\vspace{3mm}
\includegraphics[width=0.45\textwidth]{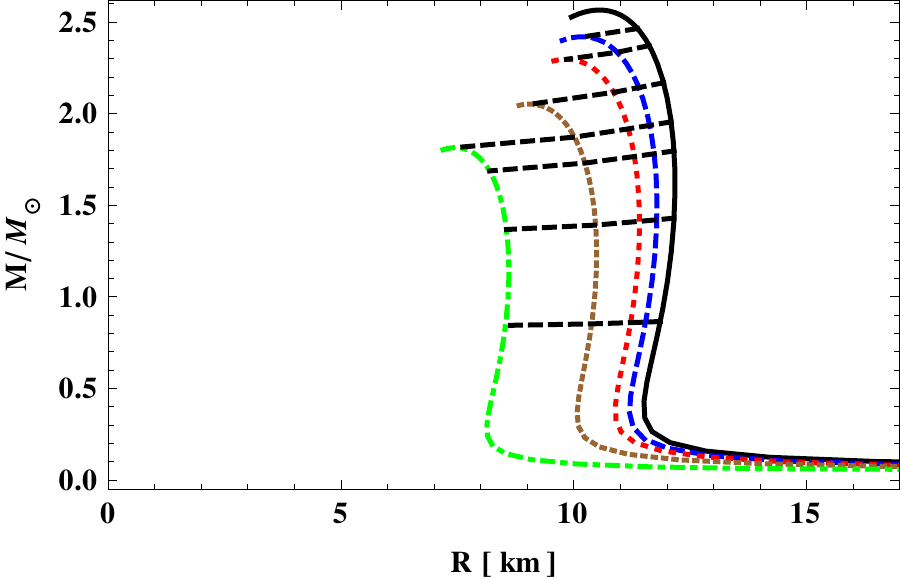}
\caption{ \label{fig:Sly}
Mass-radius diagrams obtained with two different EoS: Sly (upper) and joined polytropes 
(lower).  Black solid curves correspond to initial configurations ($\chi=0$) 
whereas blue, red, brown and green dashed curves correspond to mixed configurations 
respectively for $\chi = 0.5,0.6,0.75$ and 1. Horizontal dashed curves indicate the evolution 
of a given NS with a conserved overall baryon number $N_\text{O}+N_\text{M}=N$. 
}
\end{figure}

An important aspect is that very massive neutron stars cannot evolve in MMS,   
since they collapse to a BH when the  ratio of ordinary to mirror central densities reaches 
a critical value $\chi_\text{max}$. 
This is  illustrated in Fig. \ref{fig:Sly}, where we show the mass-radius  relations for  
a mixed star 
with different O  and M fractions    
(from $X = 0$, solid black lines, to $X=1/2$, dashed green lines) 
using the two considered EoSs. 
The nearly horizontal black dashed curves show the evolution track  with increasing values 
of $\chi$  at fixed total baryon number $N=N_\text{O} + N_\text{M}$. 
Hence, they trace the time evolution from the original NS 
($\chi=0$, solid black lines), passing the intermediate MS stage  
and eventually reaching the MMS stage ($\chi=1$, dashed green lines). 
Along the evolution tracks both the  gravitational masses and the radii decrease making 
the stellar object more compact. At a certain stage of their evolution,   very massive stars     
may become  gravitationally unstable forcing the collapse to BHs for some value of 
$\chi_{\rm max}<1$; 
the actual value of $\chi_{\rm max}$  depends on the initial NS mass and on the EoS used. 
In particular, for the Sly case only NSs with initial masses $M_\text{NS}1.55\,M_\odot$ 
or so can survive asymptotically in time and approach the MMS configuration with 
$M_\text{MMS}<1.45\,M_\odot$; 
more massive stars are doomed to collapse to BHs. 
For instance, a NS with a mass of about $1.8\,M_\odot$ collapses to BH when $\chi\simeq 0.5 $. 

\subsection{Scaling relations between the neutron stars and maximally mixed stars}
\label{sec:scaling}

The masses and radii of mixed stars  depend on 
 central densities  of the two components, $\rho_\text{cO}$ and $\rho_\text{cM}$. 
In fact, any point in the mass--radius diagram is determined by the functions
$M_\text{MS}= M(\rho_\text{cO},\rho_\text{cM})$  
and $R_\text{MS}= R(\rho_\text{cO},\rho_\text{cM})$. 
The dependence on two parameters  
causes the mass/radius degeneracy in Fig.~\ref{fig:Sly}, 
meaning that compact stars with the same mass can have different radii, depending 
on the composition.  

The masses and radii along the dashed lines in Figs.~\ref{fig:Sly},  
describing the MS evolution at fixed total baryon number $N=N_\text{O} + N_\text{M}$, 
are not related by a simple rule.  
However, one can find  simple scaling relations  between the two extreme 
configurations,   
meaning that  the mass-radius trajectories  corresponding to the MMS configurations  
(dashed green curves in Figs.~\ref{fig:Sly}) can be mapped onto those of the NS 
(black solid curves) by factor of $\sqrt 2$  rescaling of masses and radii. 
In other words, the radii and masses (as well as total  baryon numbers)  
of stars with $\rho_\text{cO}=\rho_\text{cM}=\rho_c$  are related to those of stars  
with $\rho_\text{cO}=\rho_c$, $\rho_\text{cM}=0$,  as  
\begin{align}\label{sqrt2} 
\frac{R(\rho_c,\rho_c)}{R(\rho_c,0)} = \frac{M(\rho_c,\rho_c)}{M(\rho_c,0)} = 
\frac{N(\rho_c,\rho_c)}{N(\rho_c,0)} = \frac{1}{\sqrt2}
\end{align}
The  MS configurations can be equivalently parametrized  
in terms of total central density $\rho_c= \rho_\text{cO}+\rho_\text{cM}$ 
and the ratio $\chi = \rho_\text{cM}/\rho_\text{cO}$. 
Since the  MMS and  NS configurations correspond to $\chi=1$ and $\chi=0$ respectively, 
we can consider their radii as functions of respective total central densities, denoting 
$R(\rho_c,\rho_c)\equiv R_\text{MMS}(2\rho_c)$ and $R(\rho_c,0)\equiv R_\text{NS}(\rho_c)$, 
and similarly for the masses and baryon numbers. 
Thus, in configurations related by  $\sqrt2$ scaling \eqref{sqrt2}  the central energy density 
and pressure of the MMS are twice the corresponding values  of the NS. 

The scaling relations  can be derived as follows. 
Since MMS  contains two fluids with the identical EoS and with equal central densities,  
 we can set  $\rho_\text{O} =\rho_\text{M} $, $p_\text{O}=p_\text{M}$ and $m_\text{O}=m_\text{M}$ 
 in Eqs. \eqref{eq:dmalpha} and \eqref{eq:system}.  
Thus, both components  satisfy identical differential equations:  
\begin{equation}\label{M12} 
 \frac{d m_\alpha}{d r} = 4 \pi  \rho_\alpha r^2 , \quad 
\frac{1}{\rho_\alpha+p_\alpha}\,\frac{d p_\alpha}{d r} 
= \frac{2 m_\alpha + 8 \pi p_\alpha r^3}{4 m_\alpha r -r^2} 
\end{equation} 
where $\alpha=\text{O or M}$.
By substituting 
\be{tildes}
r=\tilde{r}/\sqrt2, \qquad  m_\alpha = \tilde{m}_\alpha/\sqrt8 
\ee 
these equations can be rewritten as 
\begin{equation}\label{M12-mod} 
 \frac{d \tilde{m}_\alpha}{d \tilde{r}} = 4 \pi  \rho_\alpha \tilde{r}^2 , \quad 
\frac{1}{\rho_\alpha+p_\alpha}\,\frac{d p_\alpha}{d \tilde{r} } 
= \frac{\tilde{m}_\alpha + 4 \pi p_\alpha \tilde{r}^3}{2 \tilde{m}_\alpha\tilde{r} -\tilde{r}^2}  
\end{equation} 
which have exactly the same form as the one fluid TOV equations \eqref{TOV-NS}
for the ordinary NS. 
Therefore,  Eqs. \eqref{TOV-NS} and \eqref{M12-mod} 
should have identical solutions under the same boundary conditions. 
Namely, taking the central densities as $\rho_\text{O}(0)=\rho_\text{M}(0)=\rho_c$, we obtain 
\be{f-MMS} 
\rho_\text{O}(\tilde r) = \rho_\text{M}(\tilde r) = \rho_c f(\tilde r) 
\ee
with exactly the same shape function as $f(r)$ in Eq. \eqref{f-NS} but with the 
argument rescaled as $\tilde r = \sqrt2 r$.  The MMS radius 
corresponds to the distance at which $f(\sqrt2 r)$ vanishes. 
Thus, we obtain that the radii of the MMS with central density 
$\rho_\text{cO}+ \rho_\text{cM}=2 \rho_c$ and 
of the NS with the central density $\rho_\text{cO}=\rho_c$ are related as 
\be{R-scaling} 
R_\text{MMS}(2\rho_c)= \frac{1}{\sqrt{2}}R_\text{NS}(\rho_c) 
\ee
On the other hand, by integrating the first equation \eqref{M12-mod} and 
taking into account the redefinitions \eqref{f-MMS} we get 
$\tilde{m}_\alpha(\tilde r)=\sqrt8 m_\alpha(\tilde r) = 4\pi \int_0^{\tilde r} \rho_\alpha(r) r^2 dr$. 
Thus, for the MMS mass we have 
\begin{align}\label{M-scaling}  
M_\text{MMS}(2\rho_c) = \frac{4\pi}{\sqrt8} \int_0^\infty \!\!
\big[\rho_\text{O}(r)+ \rho_\text{M}(r)\big] r^2 dr  \nonumber \\
= \frac{4\pi \rho_c}{\sqrt2} \int_0^\infty \!\! f(r) r^2 dr 
=\frac{1}{\sqrt2} M_\text{NS}(\rho_c)
\end{align} 
where the last equality follows from \eqref{M-NS}.  
In addition, Eqs. \eqref{R-scaling} and \eqref{M-scaling} show that 
the two configurations must have the same compactness:  
\be{M/R} 
\frac{M_\text{MMS}(2\rho_c)}{R_\text{MMS}(2\rho_c)} = 
\frac{M_\text{NS}(\rho_c)}{R_\text{NS}(\rho_c)}
\ee
Analogously, for the moments of inetia one can obtain: 
\begin{align}\label{I-scaling}  
I_\text{MMS}(2\rho_c) =\frac{1}{2\sqrt2} I_\text{NS}(\rho_c)
\end{align} 

The following remark is in order for avoiding the confusion.  
The NS with $\rho_\text{cO}=\rho_c$ and the MMS with $\rho_\text{cO}=\rho_\text{cM}=\rho_c$ 
are not on the same evolutionary track with the conserved baryon number 
(dash lines in Figs. \ref{fig:Sly}). In fact, 
the baryon numbers of these configurations also obey to the scaling law: 
\be{N-scaling} 
N_\text{MMS}(2\rho_c)= 
\frac{1}{\sqrt{2}}N_\text{NS}(\rho_c)  
\ee
which can be obtained in the analogous manner, by comparing the NS baryon number  \eqref{B-NS}
with the total baryon number $N_\text{O} + N_\text{M}=2N_\text{O}$ of the MMS,   
and taking into account that upon redefinitions \eqref{tildes} we have 
$2m_\alpha/r = \tilde{m}_\alpha/\tilde{r}$
in $g_{rr}$ component of the metric tensor \eqref{g-metric}. 
By this reason also the volumes of two configurations rescale as 
$V_\text{MMS}/V_\text{NS} = (R_\text{MMS}/R_\text{NS})^3 = (1/\sqrt2)^{3}$, 
as it would occur in the case of flat space. 

An important implication is the following. If for given EoS the last stable NS 
 configuration corresponds to maximum central density $\rho_c^\text{max}$, 
 $M_\text{NS}^\text{max}= M_\text{NS}(\rho_c^\text{max})$, 
 then, for the same EoS, the last stable MMS configuration would 
 correspond to  $M_\text{MMS}^\text{max} = M_\text{MMS}(2\rho_c^\text{max})$. 
Hence,  the maximum masses (and respective baryon numbers) 
of the MMS and NS are related as 
 \be{max-M} 
 M_\text{MMS}^\text{max}   =  \frac{1}{\sqrt2}M_\text{NS}^\text{max}, \qquad  
 N_\text{MMS}^\text{max}   =  \frac{1}{\sqrt2}N_\text{NS}^\text{max} 
 \ee
For example, for the Sly EoS we have $M_{\rm NS}^{\rm max} \approx 2.05~M_\odot$  
and $N_{\rm NS}^{\rm max} \approx 2.91 \times 10^{57}$, and so  maximum mass 
and total baryon number of  the MMS must be $M_{\rm MMS}^{\rm max} \approx 1.45~M_\odot$ 
and $N_{\rm MMS}^{\rm max} \approx 2.06 \times 10^{57}$.    
Since total baryon number is conserved in the evolution,  the  
MMS configuration can be evolved only from the NS containing at most 
$N_{\rm NS} \approx 2.06 \times 10^{57}$ baryons, 
i.e. having initial mass at most $M_{\rm NS} \approx 1.55~M_\odot$   
(see corresponding evolution tracks  with conserved baryon numbers 
shown by dash lines in Fig.~\ref{fig:Sly}). 
Thus, the binding energy increases during the evolution by $0.1~M_\odot$. 
Stars with larger initial masses  at some stage of the evolution will collapse to a BH, 
before reaching the MMS limit.  E.g. star with initial mass $M_\text{NS} =1.8~M_\odot$
 ($N_\text{NS} \approx 2.5\times 10^{57}$) collapses when the ratio of  
 central densities reaches $\chi=0.5$. 
 The Rhoades-Ruffini bound $M_{\rm NS}^{\rm max} < 3.2~M_\odot$ 
implies $M_{\rm MMS}^{\rm max} = M_{\rm NS}^{\rm max}/\sqrt2  <  2.3~M_\odot$.

The above scaling relations 
 can be straightforwardly extended  to scenarios with more than one 
mirror sector: if ordinary  neutrons  have mixings with $k$ ``mirror'' neutrons 
$n'_1,n'_2,\dots n'_k$, and if  all mirror matters have the same EoS as ordinary matter
(E.g. such a scenario with $k\sim 10^{32}$ mirror sectors was discussed in Ref.~\cite{Dvali}), 
then the NS and MMS  configurations can be related by a $\sqrt{k}$--scaling rule analogous 
to~Eqs. \eqref{R-scaling}, \eqref{M-scaling} and \eqref{N-scaling}. 
In particular, for the maximum masses we would have 
$M_\text{MMS}^\text{max} = M_\text{NS}^\text{max}/\sqrt{k}$.

\section{The time evolution of mixed stars}
\label{sec:evolution}

\subsection{Neutron-mirror neutron oscillation in nuclear medium} 

The interior of neutron stars can be considered as degenerate nuclear matter 
dominantly consisting of the neutrons.\footnote{In the following,  for the sake of simplicity, 
we neglect subdominant fraction of protons and electrons whenever their role is not important.
Let us also remark that $n-n'$ conversion is ineffective in outer crust where 
the neutrons are bound in heavy nuclei. }
We assume  $n-n'$ conversion time to be much larger than the cooling time,   
so that at any stage  the temperature is much less 
than the chemical potential,  $T \ll \mu$.
 (in this section we use natural units $c=1$ and $\hbar=1$.) 
Hence, one can neglect $T$ and simply take $\mu$  as Fermi energy which in the approximation 
of ideal non-relativistic Fermi gas is ${E}_F \simeq \xi^{2/3} \times 60$~MeV, 
 where $\xi =n_\text{O}/n_{\rm S}$ is the ordinary baryon number density 
 in units of nuclear density $n_{\rm S} =0.16~{\rm fm}^{-3} = (107~{\rm MeV})^3$. 
However, this  approximation is not sufficient  for describing the observed NS 
(such ``ideal" EoS would give the maximum  NS mass $M_{\rm NS}^{\rm max}=0.7~M_\odot$ \cite{OV}), 
and  one has to take into account the (repulsive) nuclear interactions 
which stiffen the EoS of the medium. 
The coherent scatterings modify the dispersion relation of the neutron  
which can be accounted by the in-medium optical  potential  
\beq\label{eq:Vn}
V_n=2\pi a \, \frac{n_\text{O}}{m_n} = \xi \, a_3 \times 125\,\text{MeV} 
\eeq 
where $a$ is the neutron scattering length and $a_3=(a/3~\text{fm})$.   
At supra-nuclear densities ($\xi > 1$) 
both $V_n$ and  $E_F$ are $\sim100$~MeV, but  $V_n> E_F$. 

The oscillation $n-n'$ in nuclear medium 
is described by the Schr\"odinger equation with effective Hamiltonian  
\be{H-osc}
H = \mat{ E_n  }{\eps} {\eps} { E'_n   } 
\ee
where $E_n=(m_n^2 + p^2)^{1/2}+V_n$,  with $p$ being the neutron momentum.\footnote{
For simplicity, we consider the exact mirror model with $m_{n'}=m_n$, 
however our results are applicable as well in the case of small mass splitting. 
We also neglect the  Zeeman energy induced by the neutron magnetic moment $\mu_n$  
 since $\vert \mu_n B \vert \ll 1$~MeV 
 even for magnetars where the magnetic field $B$ can reach $10^{15}$~G or so. 
 }
 Analogously, we take  $E'_n$ with $V_n \to V'_n$,  
where $V'_n$ is the mirror neutron potential 
obtained from Eq. \eqref{eq:Vn} by substituting $n_\text{O}$ by the mirror baryon density 
$n_\text{M}$. 
Hence,  we have $V'_n=\beta V_n$ where $\beta = n_\text{M}/n_\text{O}$.  
%
The off-diagonal term $\eps$ comes from $n-n'$ mass mixing \eqref{n-npr}. 
The eigenstates  of Hamiltonian \eqref{H-osc} are 
\begin{align} n_1 = c\, n - s\, n'   \, , \quad\quad 
n_2  = s\, n  +  c\, n'   
\label{eq:eigen} 
\end{align} 
where $c= \cos \theta$, $s=  \sin\theta$, and $\theta$ is the mixing angle:
\be{arctan}
\tan 2\theta=2\eps/\Delta\! E       
\ee 
where  $\Delta\! E =  E_n - E'_n = V_n(1-\beta)$.    
%
The probability of $n-n'$ oscillation  for flight time $t$ is  
\beqn{Pnn} 
P_{nn'}  =  \sin^2 2\theta \sin^2(\Delta\! E\, t/2)  
\eeqn  
For $\Delta\! E\, t \ll 1$ we have $P_{nn'} = (\eps t)^2$,  while for 
$\Delta\! E\, t  \gg 1$ the 
time dependence can be averaged so that $P_{nn'} = 
\frac12 \sin^2 2\theta$.  
This can be simply interpreted also in following way. Creating the neutron as initial state 
is equivalent of creating  the eigenstates $n_1$ and $n_2$
with the probabilities $c^2$ and $s^2$ respectively. The eigenstates do not oscillate 
among each other, and after freely propagating to some distance,   
one can detect $n_1$ and $n_2$ as $n'$  with probabilities 
$s^2$ and $c^2$  correspondingly. 
Thus, by combining the probabilities, we get $P_{nn'} = 2 c^2s^2 = \frac12 \sin^2 2\theta$. 

Once again, if the neutrons are considered as ideal (non-interacting) gas, 
$n-n'$ oscillation per s\`e cannot lead to the neutron star 
transformation into the mixed star.  
As far as $\Delta\!E \gg \eps$, the mixing is extremely small:\footnote{In real situations  
the condition $\Delta\!E \gg \eps$ 
is always satisfied  as far as $\eps$ is very small. In the limit  $\Delta E\to 0$ 
(i.e. $\beta=0$)  $\theta$ has no singularity since Eq. \eqref{arctan} implies 
maximal mixing $\theta=\pi/4$.} 
\be{theta}
 \theta = \frac{\eps}{\Delta E}  = 
 \frac{\eps_{15}}{\xi(1-\beta)a_3 } \times 8\cdot 10^{-24} 
\ee
where $\eps_{15} = (\eps/10^{-15}\,\rm eV)$. 
Hence, for the freely propagating neutrons  
the M neutron fraction remains negligibly small at any time,  
$P_{nn'} \simeq 2\theta^2\ll 1$.  Therefore, one has to take into 
account the neutron interactions in the medium. 

\subsection{Neutron--mirror neutron conversion processes } 

Although the ordinary and mirror nucleons have separate strong interactions, 
the mixed interactions emerge in the basis of the Hamiltonian eigenstates \eqref{eq:eigen}.  
E.g.  the  couplings 
$\pi^0 \,\ov{n} \gamma^5 n + \pi^{0\prime}\, \ov{n'} \gamma^5 n' $ 
with ordinary and mirror neutral pions (the coupling constant $g_{\pi NN}$ is omitted)
%
give rise to non-diagonal terms between
 $n_1$ and $n_2$ states: 
 \be{piNN} 
s \,  \big(\pi^0 \, \ov{n_2} \gamma^5 n_1 + \pi^{0\prime} \, \ov{n_2} \gamma^5 n_1\big) + \text{h.c.} 
\ee
The same occurs for the analogous couplings with $\rho$ mesons, etc. 
More generically,  
 strong interactions  of the neutron with the target nucleons ($N=n,p$) 
 via single or multi-pion and $\rho$-meson exchanges etc.  
 can be described by  the effective couplings $(\ov{n} \lambda n)(\ov{N}\lambda N)$
 where $\lambda=\gamma^5, \gamma^\mu$ etc. stand for the possible Lorentz structures.   
Since  
$n= c\, n_1 + s\, n_2$, the above interactions 
 in the  basis of eigenstates  have the form 
\be{non-diag}
\big(c^2\, \ov{n_1} \lambda n_1+s^2 \, \ov{n_2} \lambda n_2 
+cs\, \ov{n_2} \lambda n_1 + cs\, \ov{n_1} \lambda n_2 \big)\big(\ov{N}\lambda N\big)
\ee
which contains also mixed entries between $n_1$ and $n_2$.  

\begin{figure}
\includegraphics[width=0.4\textwidth]{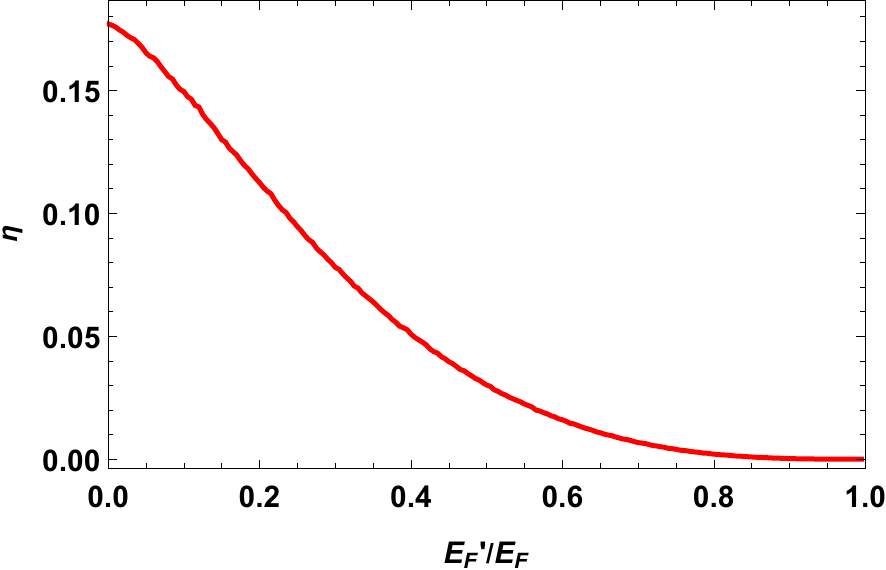}
\caption{
Pauli blocking factor $\eta$ as a function of $E'_F/E_F$. 
}
\label{fig:eta}
\end{figure}

The mirror neutron production rate can be estimated as follows. 
The stationary neutron state in the nuclear medium 
can be viewed as the eigenstate  $n_1$  
(barring a tiny ($\approx \theta^2$) fraction of $n_2$).  
The processes $n_1N \to n_1 N$, $N=n,p$  
are Pauli blocked,  but $n_1N\to n_2 N$ are not blocked.  
Therefore, every $n_1n_1$ collision can produce $n_2 \approx n'$ with 
a cross section $2\theta^2 \sigma_{nn}$ where 
$\sigma = 4\pi a^2$ is $nn$ scattering cross section.\footnote{The process $n_1n_1\to n_2 n_2$ 
has no Pauli suppression but its rate is proportional to $\theta^4$ and thus it is negligible. 
The processes $n_1 p \to n_2 p$ involving protons and heavier baryons 
contribute with the rate $\propto \theta^2$, but for simplicity we neglect them 
because of small ($\sim 10\,\%$) fraction of protons in the NS interior.  }
Taking  the mean  relative velocity as $v=p_F/m_n$, 
 the rate of $n \to n'$ transformation can be estimated as 
 \be{Gamma} 
 \Gamma(\xi,\beta) = 2\theta^2 \eta  \langle \sigma v\rangle \, n_\text{O}  
= \frac{2 \eta(\beta) \eps^2 m_n p_F}{\pi (1-\beta)^2 \xi n_\text{S} } 
 \ee
where for $\theta$ we take Eq. \eqref{theta} 
with $\Delta E = V_n - V'_n = 2\pi a (1-\beta)n_\text{O}/m_n$. Thus, the dependence 
on the poorly known scattering length $a$ in fact cancels out in Eq. \eqref{Gamma}.  
The Pauli blocking factor $\eta$ takes into account that the final state $n_1$ 
cannot have a momentum below the Fermi momentum 
$p_F = (2m_nE_F)^{1/2} \simeq \xi^{1/3} \times 340$~MeV  
while the momentum of the produced $n_2 \approx n'$  state should be above 
$p'_F\simeq (\beta\xi)^{1/3} \times 340$~MeV.  
The blocking factor $\eta(\beta)$ as a function of $E'_F/E_F=\beta^{2/3}$
is estimated  by a Montecarlo  simulation of hard sphere  scatterings,  
and the resulting dependence  is reported in Fig. \ref{fig:eta}. 
This function has 
 a maximum $\eta(0) \approx 0.18$ at $E'_F=0$,  then it  decreases 
and, as expected, asymptotically vanishes when the star  approaches the MMS  state, 
$E'_F \to E_F$.\footnote{Let us remark that  the reverse process $n_2n_2 \to n_2 n_1$
 is Pauli blocked as far as $E_F > E'_F$.   
This justifies why we have neglected the inverse reaction rate $\Gamma'$ in  Eq.~\eqref{Boltzmann}
to obtain Eq.~\eqref{Boltzmann-X}.
} 
For the younger MS, with $n_\text{M} \ll n_\text{O}$, we can set $\beta=0$ 
and  Eq. \eqref{Gamma} gives 
 \be{Gamma-0} 
\Gamma(\xi,0)  = \eps_{15}^2 \xi^{-2/3}  
\times 3.0 \cdot 10^{-47} ~ \text{GeV}
 \ee 
This rate depends on the baryon density, meaning that $n-n'$ conversion should proceed 
somewhat  faster in peripheral low density regions of the star 
rather than in central regions where $\xi > 1$. 

 Since the process $n_1n_1 \to n_1 n_2$ (hereafter we call it simply as $nn \to nn'$ process) 
 takes place at the momenta 
close to the Fermi surface, the mirror neutron is produced with typical energy
$E_F = \xi^{2/3} \times 60$~MeV.  
This energy will be radiated away by cooling and produced mirror neutrons 
start to form  degenerate core with $E'_F < E_F$. 
(In fact, mirror neutrons produced at very initial stages will decay as $n' \to p'e'\bar\nu'$ 
and cooling can be related to $p'$ and $e'$ components 
which then undergo the `neutronization' in the core.)
Therefore, the energy production rate per baryon  
can be roughly estimated as 
\be{Gamma-E} 
\Gamma(\xi,0)  E_F(\xi) \simeq \eps_{15}^2  \times 2.7\cdot 10^{-24} ~{\rm GeV/s}
\ee
Notice that the $\xi$-dependence is cancelled.

Mirror neutrons can be produced also by other processes. 
In normal conditions the mixed interaction terms \eqref{piNN} 
cannot induce decays $n_1 \to n_2 \pi (\pi^{\prime})$ with the emission of 
neutral (ordinary and mirror) pions, simply by kinematical reason.   
However, such decays become possible in dense nuclear medium. 
This effect is similar to the matter induced neutrino decay  \cite{Matter-nu}, 
and we shall simply call it $n\to n'$ decay.    
Namely,  considering the large enough densities at which the threshold condition 
$\Delta E= V_n  > m_\pi$ is satisfied,  
the matter-induced decays $n_1 \to n_2 \pi^0(\pi^{0\prime})$ 
can proceed with no Pauli blocking.  The decay rates (taking again $\beta=0$) read  
\begin{align}\label{gpiNN} 
 \Gamma_{n\to n' } 
& = \frac{\theta^2 g_{\pi N}^2  (\Delta E - m_\pi)^3 }{8\pi m_n^2} 
= \frac{\eps^2 g_A^2 V_n }{8 \pi F_\pi^2} \left(1-\frac{m_\pi}{V_n}\right)^3
\nonumber \\ 
& 
\simeq  \eps_{15}^2 (\xi a_3-1.1)^3/(\xi a_3)^{2}  \times 8 \cdot  10^{-49} ~\text{GeV} 
\end{align}
where the Goldberger-Treiman relation  
$g_{\pi N}= g_A m_n/F_\pi $ is used for the pion-nucleon coupling constant, 
with $F_\pi=93$~MeV being the pion decay constant and $g_A=1.27$  
the axial coupling constant. 
The decay $n \to p' \pi^{\prime}$ into mirror proton and  "negative" mirror pion 
is also possible.  
However, the rate \eqref{gpiNN} can be competitive 
with \eqref{Gamma-0} only in central regions of the heavy NS 
where the density can be as large as $\xi =8$ or so. 
Therefore, in the following our estimations 
will be based on the rate given by Eq. \eqref{Gamma-0}. 

The heavy eigenstate can also decay into the lighter one  with emission of 
the ordinary or mirror photons, $n_1\to n_2 \gamma(\gamma')$, 
via the transitional magnetic moment (TMM) $\mu_{12}$ between the two eigenstates.
The decay rates read  
\be{mu12} 
\Gamma_{n\to n'\gamma} = \Gamma_{n\to n'\gamma'} = \frac{\mu_{12}^2}{\pi}  \Delta E^3 
\ee
In particular, the mass mixing \eqref{n-npr} induces the TMM $\mu_{12} = \theta \mu_n$ 
between $n_1$ and $n_2$ states right from the magnetic moments  of $n$ and $n'$, 
$\mu_n= \mu'_n=-1.91 \mu_N$, where $\mu_N = e/2m_N$ is the nuclear magneton. 
Then we get
\be{mu12-theta} 
\Gamma_{n\to n'\gamma'(\gamma)}  = \frac{\eps^2\mu_{n}^2}{\pi}  V_n  
= \eps_{15}^2 a_3 \xi \times 3\cdot 10^{-51}~\text{GeV}
\ee
 which is four orders of magnitude below \eqref{Gamma-0}.
 Hence,  $n\to n'\gamma$ decay cannot be the dominant effect if the TMM 
 is induced solely by $n-n'$ mass mixing \eqref{n-npr}.  
 However, 
 the situation changes if there is a direct TMM 
 $\mu_{nn'}$  between ordinary and mirror neutrons, in which case 
 we would have $\mu_{12} = \mu_{nn'}$.  
 In fact, while the TMM between the neutron and antineutron 
 is excluded by fundamental symmetry reasons, it is allowed between $n$ and $n'$ \cite{Arkady}, 
 and can be induced in some models of $n-n'$ mixing \cite{LHEP}. 
 Present experimental limits imply the upper bound $\mu_{nn'} < 10^{-5} \mu_n$ or so \cite{MDPI}. 
Clearly,  for the large enough TMM, $\mu_{nn'} \gg  \theta \mu_{n}$, the rate \eqref{mu12} 
can be larger than \eqref{Gamma-0}. 
However, here we shall not discuss this situation. 

The role of weak interactions is negligible. 
The Lagrangian describing mirror neutron $\beta$-decay $n' \to p'e'\bar{\nu}'_e$, 
 induces  $n_1 \to p'e'\bar{\nu}'_e$ decay via the admixture of $n_1$  in $n'$.
It produces mirror protons and electrons with the rate 
\begin{align} \label{Gamma-weak} 
  \frac{\theta^2 G_F^2(1+ 3g_A^2) \Delta E^5 }{60 \pi^3} 
= \eps_{15}^2 (a_3 \xi)^3 \times 4 \cdot 10^{-64} ~\text{GeV} 
\end{align}
which is vanishingly small as compared to \eqref{Gamma-0}.

 \subsection{Estimating the evolution time}
 
  
Let us consider the evolution track of a star with a given overall baryon number 
$N=N_\text{O}+N_\text{M}$, starting from its initial (NS) configuration with $N_\text{O}=N$ 
and $N_\text{M}=0$. 
Since $n-n'$ transition rate \eqref{Gamma} depends on $\xi$ and $\beta$,  
these values should be averaged over the density distributions in the star. 
As far as the adiabatic evolution is assumed, at any time $t$ the profiles of the number densities 
$n_\text{O}(r) = \xi(r) n_\text{S}$ and $n_\text{M}(r) = \beta(r)\xi(r) n_\text{S}$, 
as the functions of radial coordinate $r$,   
are fully determined by the mirror baryon fraction $X(t)=N_\text{M}(t)/N$.  
Hence, we have  
\be{Gamma-X} 
\Gamma(X) = \Gamma_0 \cF(X) 
\ee 
where 
\be{F-X} 
\cF(X)=    \frac{ \langle \xi_0 \rangle^{2/3} } { \eta(0) }  
\left\langle \frac{  \eta(\beta_X)  } { \xi_X^{2/3} (1-\beta_X)^2 }\right\rangle 
\ee 
 with the parentheses meaning the average  
 over the density distributions in the star at given M fraction $X$, 
and 
 \be{Gamma-X0} 
\Gamma_0  = (\dot{N}/N)_{X=0} = \eps^2_{15}\langle \xi_0^{-2/3} \rangle  
\times 4.56 \cdot 10^{-23} ~ \text{s}^{-1}
 \ee 
 is  the `starting' conversion  rate  obtained  from Eq. \eqref{Gamma-0} by averaging 
 over $\xi(r)$ distribution in the initial NS  with $X=0$ (i.e. $\beta=0$).   
Considering a  NS containing $N$ baryons and having a radius $R$, 
we have $\langle n_\text{O} \rangle_{X=0} = N/V$ where $V \approx  (4\pi/3) R^3$
(neglecting the few per cent GR corrections for the NS volume).  
Thus we get  $\langle \xi_0 \rangle \approx R_{12}^{-3}(N/N_\odot)$ 
where $R_{12}=R/12\,\text{km}$ and $N_{\odot} = 1.188\times 10^{57}$. 
Then one can define characteristic conversion time $\tau_\eps$ as 
the inverse value of $\Gamma_0$: 
\begin{align}\label{tau-eps} 
 \tau_\eps = \Gamma_0^{-1} & \simeq 
\eps_{15}^{-2} \, R_{12}^{-2} \, (N/N_\odot)^{2/3}   \times 0.7 \cdot 10^{15} ~{\rm yr}  \nonumber \\
& \simeq \eps_{15}^{-2} \, R_{12}^{-2} \, (\kappa M/1.5\, M_\odot)^{2/3}   \times 10^{15} ~{\rm yr} 
\end{align} 
where for the last step we used the relation \eqref{M-R-Sly}. 
For a given EoS, the total baryon number determines it the mass and radius. 
For massive stars with $M \simeq 2\, M_\odot$ the Sly EoS  implies $\kappa =1.2$
ands $R=10$~km, and  
the respective evolution time 
can be estimated as 
\be{tau-Sly}
\tau_\eps^\text{Sly} (2M_\odot) \approx  \eps_{15}^{-12}  \times 2 \cdot 10^{15} ~{\rm yr} 
\ee
Other EoS can give somewhat different results, but within factor of two or so.  


Integrating  Eq. (\ref{Boltzmann-X})  with $\Gamma$ given by \eqref{Gamma-X}, 
we obtain the age of the MS in which the mirror baryon fraction has reached a value $X<1/2$:  
\be{t}
t(X)=  \int_0^{X}\!\!\! \frac{dx}{\Gamma(x) (1-x) } =  \tau_\eps \int_0^{X}\! \!\frac{dx }{(1-x)\cF(x) } 
\ee
Since $\cF \to 1$ when $x \to 0$,  
for younger MS 
we obtain 
 \be{t-smallX}
 X(t) = t/\tau_\eps
  \ee
 meaning that $X$  linearly increases in  time 
  until it remains small enough, $X \ll 1$. But the evolution gradually slows down 
  with growing $X$, since  the transformation rate  \eqref{F-X} 
  decreases due to increase of $\xi$ and $\beta$ and respective decrease of Pauli factor $\eta(\beta)$. 
  
 During the evolution  the ratio of the central densities $\chi=\rho_\text{cM}/\rho_\text{cO}$
 increases along with the M fraction $X$. 
For the lighter stars that can reach the MMS stage with 
with $\chi =1$,  Eq. \eqref{t} can be integrated up to $X=1/2$. 
 
However,  the heavier MS at a certain moment 
should collapse into BHs and so Eq.~\eqref{t} should be integrated up to some value 
$X<1/2$ at which $\chi(X)$ reaches the critical value $\chi_\text{max}$,  
which in turn depends on the initial mass of the progenitor NS as well as on the EoS. 
 For example, for the case of the Sly EoS 
the NS with initial mass $M \approx1.8~M_\odot$ will collapse at $\chi\approx 0.5$ 
(with its gravitational mass reduced to $1.7~M_\odot$) 
which corresponds to the evolution time $t \approx 0.1 \tau_{\eps}$.  
The stars with smaller mass can have larger lifetimes: e.g. a NS born with 
$M =1.6~M_\odot$ can survive for time $t \simeq \tau_\eps$, 
and will collapse with mass reduced to $1.5~M_\odot$ or so.  
And finally,  stars with initial masses $M \leq 1.55~M_\odot$ 
do not collapse, and asymptotically in time can reach the MMS configuration 
with masses rescaled down to $M \leq 1.45~M_\odot$, gaining the gravitational 
binding energy of about $0.1~M_\odot$.  

\begin{figure}
\includegraphics[width=0.40\textwidth]{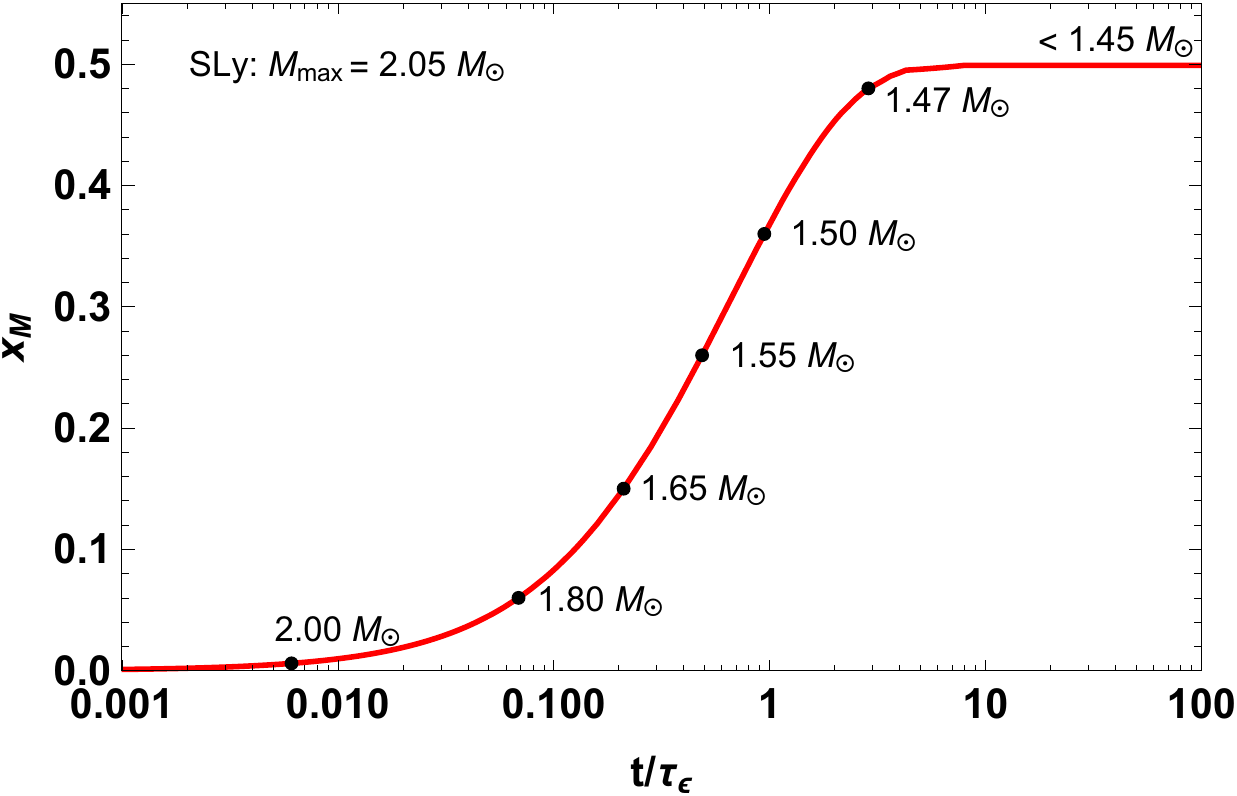}
\includegraphics[width=0.40\textwidth]{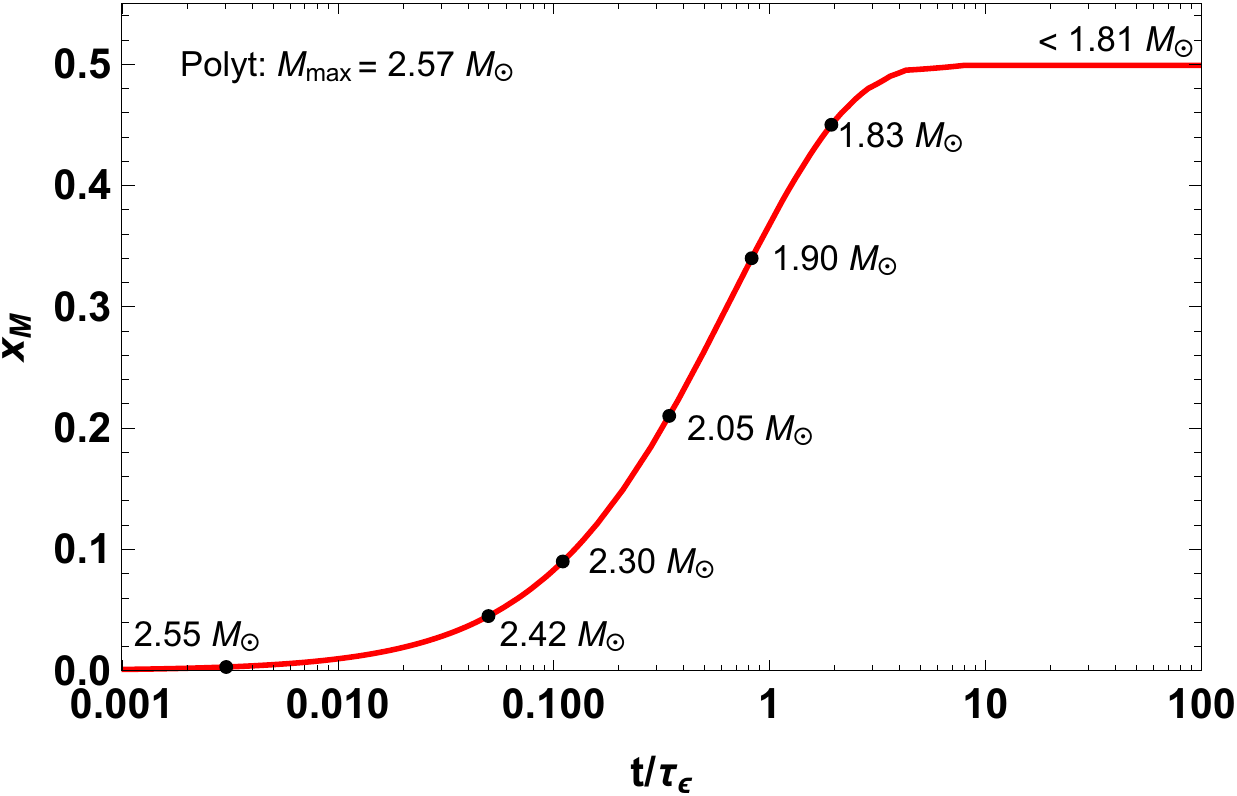}
\caption{Evolution of the mirror fraction $X=N_\text{M}/N$ in  mixed stars as a function of time
in units of  $\tau_\eps = \Gamma_0^{-1}$. 
The black points on the solid curves show the maximal possible mass of the MS
at the corresponding age. }
\label{fig:mass-time}
\end{figure}

Fig. \ref{fig:mass-time} shows the time evolution of the mirror fraction $X$ 
in a MS in units of the characteristic time highlighting the maximal possible mass 
of a MS at the age $t$, after which time they collapse.  
The curves on Fig. \ref{fig:mass-time}, corresponding to our two examples of EoS,  
Sly and joined polytrope, are obtained by numerical calculations of factors $\cF(X)$ \eqref{F-X}
by averaging over the density profiles of mixed stars containing a fraction $X$ of mirror baryons 
and respective values of $\chi(X)$.


The ``starting" rate of the NS energy loss due to $n-n'$ conversion is obtained by multiplying the 
energy production rate per baryon \eqref{Gamma-E} on the baryon amount  \eqref{M-R-Sly}: 
\be{Edot} 
\dot E_{nn'} = - \eps_{15}^2 (\kappa M/1.5\,M_\odot) \times 8 \cdot 10^{30}~\text{erg/s} 
\ee 
which is applicable for the evolution times $t \ll \tau_\eps$. For larger $t$ becoming comparable 
with $\tau_\eps$ the energy loss rate decreases. 
Eq. \eqref{Edot} can be compared to the energy radiation rate due to the pulsar 
magnetic dipole field 
\be{E-magn}
\dot{E}_\text{magn} = - \frac16 B_\alpha^2 R^6 \Omega^4 \simeq 
B_9^2 R_{12}^6 P_{10}^{-4} \times 3\cdot 10^{33}~\text{erg/s}
\ee
Here $\Omega=2\pi/P$ is the pulsar rotation angular frequency ($P_{10}=P/10\,\text{ms}$) 
and $B_\alpha = B\sin\alpha$, where 
$B$ is the magnetic field strength at the pole and $\alpha$ is the angle between 
the pulsar magnetic moment and its rotation axis ($B_9=B_\alpha/10^9\,\text{G}$).  
Hence, for $\eps=10^{-15}$~eV or so, for the typical pulsars   
one expects that $\dot E_{nn'} \ll \dot{E}_\text{magn}$.  
But  for larger values of $\eps$ the two rates can be comparable and one can even have 
$ \dot E_{nn'} > \dot{E}_\text{magn}$. Let us note that  $\dot E_{nn'}$ \eqref{Edot} is 
a rather regular quantity essentially depending only on the inferred value of $\eps$, 
while $\dot{E}_\text{magn}$ \eqref{E-magn} depends on the pulsar magnetic field 
and its rotation frequency and thus can vary by orders of magnitude between different pulsars, 
achieving very large values for magnetars.


\section{Comparison with astrophysical observations}
\label{sec:pulsars}

Let us discuss now the possible astrophysical implications of $n-n'$ conversion in 
the neutron stars and derive limits on  the transformation time $\tau_\eps$ \eqref{tau-eps}
which in turn can be translated into the bounds on $\eps$.  
Most of these effects are model dependent,  and one should be careful in their interpretation. 
Namely, they can depend on the EoS as well as on the specific environmental conditions for some stars. 
For the sake of simplicity, we shall discuss two possible situations. 
First we discuss the case  
when the transformation time $\tau_\eps$ 
is larger than the age of the universe $t_U = 14$~Gyr.  
In this case no star can reach the MMS configurations: all NS should  still be neutron dominated  
hosting  in in their interior only a small mirror fraction proportional to their age $t$, $X = t/\tau_\eps \ll 1$. 
We shall discuss effects of $n-n'$ transformation for the timing, mass loss and heating of pulsars. 
Then we shall concentrate on the possibility of $\tau_\eps$  being much smaller  
than $t_U$. In this case all stars older than $\tau_\eps$ should be already transformed into the MMSs,   
 and thus should be more compact than the younger stars.


\subsection{Effects of slow NS to MS transformation}

The pulsar ages can be estimated assuming that the spin-down rate of the pulsar rotation 
 is dominated by  the energy radiated by  a rotating dipole magnetic moment. 
 If this is the only braking mechanism, the  pulsar age $t$  is given by the relation 
\begin{equation}\label{p}
t = \frac{P(t)}{2 \dot{P}(t)} \left[1 - \left(\frac{P(0)}{P(t)}\right)^2 \right] 
\end{equation} 
where $P(t)$ is the rotation period at the time $t$ and 
$P(0)$ is the rotation period at the NS birth, $t=0$. 
The measurable value $\tau_c = P/(2\dot{P})= - \Omega/(2\dot\Omega)$,  
usually called the pulsar characteristic  (or spin-down) age, 
coincides with the true age if $P(0) \ll P(t)$, meaning that the pulsar rotation 
period at birth  was much smaller than  at present time, 
and if the magnetic dipole emission has  always been dominant over other slowing down processes. 
For most of known pulsars, more prominently for those which were observed 
with large  spin-down rates $\dot{P}/P$, the true age $t$ can be rather close to 
their characteristic age $\tau_c$.  

The heaviest recycled pulsars observed up to now and their characteristic ages  
are listed in Table \ref{TablePSR}. 
The masses of first three pulsars in this Table  are compatible with $2\, M_\odot$,  
(within $1.5\sigma$ error bars for PSR J0740+6620). 
Thus in the context of the Sly EoS, which admits  the standard NS 
with masses up to  $M^{\rm max}_{\chi=0} \approx 2.05 M_\odot$, 
these pulsars would collapse due to $n-n'$ within the time $t < \tau_\eps/200$ or so 
as its is shown on upper Fig. \ref{fig:mass-time}. 
All these pulsars have spin-down ages $\tau_c$ of few Gyr. 
Thus, assuming that their true ages correspond to the respective $\tau_c$ values,  
the very existence of these pulsars  in the context of the Sly EoS  
would imply a conservative limit $\tau_\eps > 200 \tau_c \sim 10^{12}$~yr or so.
Using Eq. \eqref{tau-Sly}, this in turn can be translated into the upper bound 
$\eps < 2 \times 10^{-14}$~eV or so. 
  
The case of the globular cluster pulsar J1748$-$2021B (or NGC 6440B) 
is  more interesting. 
In spite of large errors  in the mass determination: $M =( 2.74 \pm 0.21)\, M_\odot$ ~\cite{Freire},  
already the existence of this pulsar challenges the SLy EoS 
and demands a rather stiff EoS as in Refs. \cite{Lattimer,Mueller}.\footnote{Though, 
conservatively thinking,  there is $1\%$ probability of  low enough inclination 
allowing the pulsar mass below $2M_\odot$~\cite{Freire}, so     
that it can be marginally reconciled also with the Sly EoS.}  
Namely, within the $1.5\sigma$ error, we have $M > 2.42\, M_\odot$ 
which is well compatible with our second example, the piecewise polytrope. 
Now, from lower Fig. \ref{fig:mass-time} we see that with this mass the star had to collapse 
within the time $t=\tau_\eps/20$ or so. 
But the  true age of this object cannot be determined 
since the measured value of its $\dot{P}$ is negative: instead of spinning-down, 
this pulsar seemingly spins-up.


\begin{table}
\centering
\begin{tabular}{l @{\hspace{3\tabcolsep}}  c @{\hspace{4\tabcolsep}}  c @{\hspace{4\tabcolsep}}  c 
@{\hspace{4\tabcolsep}} c @{\hspace{4\tabcolsep}} c }
\hline
\quad PSR  &  $M$ [$M_\odot$]  & $P$ [ms] & $\dot{P}[10^{-20}]$ & $\tau_c$ [Gyr]  &    \\
\hline
J1614$-$2230  \cite{Demorest} & $1.97(04)$ & $3.1508$ & ~0.962  & $5.2$ &   \\
J0348+0432 \cite{Antoniadis} &  $2.01(04)$  &  $39.123$ & ~24.07 & $2.6$ & \\
J0740+6620 \cite{Cromartie} &  $2.14(09)$  &   $2.8857$ & ~1.219 &  $3.8$ & \\
J1748-2021B  \cite{Freire}     & $2.74(21)$ &   $16.760$ & $-32.91$ &  $-1.1$ ? &  \\  
\hline
\end{tabular}
\caption{\label{TablePSR} 
Some of most massive pulsars, with their measured $P$ and $\dot{P}$ values and 
derived characteristic ages $\tau_c = P/2\dot{P}$. 
}
\end{table}

As we discuss below, in the presence of $n-n'$ conversion 
the NS true age and its characteristic age are no more related via Eq. \eqref{p}. 
 In fact, this equation 
 follows from integration of the pulsar evolution differential equation 
\be{PSR-diff}
\dot{\Omega} = - \frac{ B_\alpha^2 R^6}{6 I} \, \Omega^3   
\ee
which is obtained by equating the energy radiation rate \eqref{E-magn} 
with the time derivative of the pulsar kinetic rotational energy $E_\text{rot} = I\Omega^2/2$  
and assuming that the moment of inertia $I$ remains constant in time, 
so that $\dot{E}_\text{rot} = I\Omega \dot{\Omega}$.  
Then, using the measured values of $\Omega$ and $\dot{\Omega}$,  
 from this equation one can also determine the value 
of the pulsar magnetic field and the respective rate of the pulsar energy loss \eqref{E-magn}. 
Let us remark, however, that these are {\it derived} parameters obtained under 
assumption that the pulsar losses the rotational energy  
dominantly by the radiation due to its magnetic dipole. 

However, in the presence of $n-n'$ conversion the moment of inertia is not constant,  
and we have $\dot{E}_\text{rot} = I\Omega \dot{\Omega} + \dot{I} \Omega^2/2$.  
Hence, 
the pulsar evolution differential equation should be modified as 
\be{PSR-diff-1}
\dot{\Omega} = - \frac{ B_\alpha^2 R^6}{6 I} \, \Omega^3 - \frac{\dot{I}}{2I}\,  \Omega 
\ee
The value $\dot{I}/I= \dot{M}/M + 2\dot{R}/R$ is negative 
as far as  in the process of $n-n'$ transformation 
both the mass and radius of the star decrease. 
If this term 
is significantly large, then Eq.~\eqref{p} becomes invalid 
and the pulsar age $t$ cannot be related  to its spin-down time $\tau_c$.   
In fact, the heavy pulsars in Table \ref{TablePSR} 
could have small spin-down rates not because they are very old and posses very small magnetic 
fields,\footnote{E.g., for millisecond pulsar J1614$-$2230
the {\it derived} values of surface magnetic field and spin-down luminosity were 
evaluated respectively as $B\simeq 2\times 10^8$~G and 
$\dot{E}_\text{magn} \simeq  10^{34}$~erg/s \cite{Demorest}.}  
but due to partial cancellation between the first (negative) and 
second (positive) terms in \eqref{PSR-diff-1}. 
In this way, the actual values of magnetic fields can be larger than the derived ones, 
and the true ages of these pulsars can be much less than their spin-down ages. 
Therefore, the naive limits on $n-n'$ mixing, as $\eps < 3\times 10^{-14}$~eV obtained 
by estimating the collapse time of $2\,M_\odot$ stars in the context 
of the Sly EoS, are not applicable. 

Moreover, if in Eq. \eqref{PSR-diff-1} the positive contribution $-(\dot{I}/2I) \,\Omega$ is dominant, 
then we would have $\dot{\Omega} > 0$ which means that the pulsar spins-up. 
In particular, it is tempting to propose that the negative $\dot{P}$ of 
PSR J1748$-$2021B in Table \ref{TablePSR} is originated from this effect,  
meaning that this pulsar is spinning-up instead of slowing down 
(though this effect could be related also to the gravitational acceleration in globular cluster). 
Interestingly,  in NGC6440 and NGC6441 there are other pulsars
with $\dot P < 0$, two isolated ones as J1748$-$2021C and J1750$-$3703C, and 
one with a light companion, J1748$-$2021F \cite{Freire}.  
If future observations will find some other spinning-up pulsars, with the 
intrinsic value of $\dot{P}/P$ being confidently negative after subtracting the acceleration effects, 
then this could be interpreted as the effect of the star contraction 
due to $n-n'$ transformation. 



The interesting possibility that the neutron star mass loss 
during its transformation into mixed star can affect the orbital period 
 in the the NS binaries was discussed in Ref.~\cite{Nussinov-new}. 
 If the system losses the mass, the orbital period $P_b$  should increases 
with the rate which  is related to the mass loss rate: 
\be{Pdot} 
\frac{\dot{P}_b^\text{nn'}}{P_b} = - \frac{2 (\dot{M} + \dot{M}_c)}{M + M_c} 
\ee
where $M$ is the NS mass and $M_c$ is the mass of its companion in the binary system
which can be e.g. a white dwarf of another neutron star.  

 Assuming again that $\tau_\eps > t_U$ and using Eq. \eqref{Edot}, 
 the mass loss rate by the neutron star can be estimated as
\be{Mdot} 
\frac{\dot{M}}{M}= \frac{\dot{E}_{nn'}}{M} = - \eps_{15}^2 \kappa \times 0.9 \cdot 10^{-16}~\text{yr}^{-1}
\ee
In this way, the observational data on the orbital period decay $\dot{P}_b/P_b$ 
can be used to obtain the limits on  $\eps$. 

In fact, several pulsar binaries have positive measured values of $\dot{P}_b$. 
However, apart of positive contribution of $n-n'$ effect, the observed value  
should be corrected for the dynamical effects related  to the system acceleration 
in the galactic gravitational potential, kinematic Shklovskii effect of apparent acceleration 
and, for compact systems, also for the quadrupolar emission of gravitational waves (GW): 
$\dot{P}_b=\dot{P}^{nn'}_b  + \dot{P}_b^\text{gal} + \dot{P}_b^\text{kin} + \dot{P}_b^\text{GW}$. 
For example, the binary system containing one of the most stable pulsars J0437$-$4715,  
with mass $M\approx 1.4\,M_\odot$ and companion mass $M_c\approx 0.25\,M_\odot$, 
one has $P_b=5.7410$~days and $\dot{P}_b^\text{obs}= (3.73 \pm 0.06)\times 10^{-12}$ 
\cite{Verbiest:2008}.  
However, after subtracting the above effects,  also taking into account that the GW emission   
is negligible for this system,  one obtains 
\be{Verbiest}
\frac{\dot{P}^{nn'}_b}{P_b} = \frac{\dot{P}_b - \dot{P}_b^\text{gal} - \dot{P}_b^\text{kin}}{P_b}
=(1.0 \pm 1.8) \times 10^{-11}/\text{yr}
\ee 
which is compatible with vanishing effect. Therefore, within $1\sigma$ error, we have 
$\dot{P}_b/P_b < 2.8 \times 10^{-11}/\text{yr}$. Thus, assuming $\dot{M}_c=0$ 
(meaning that the companion mass loss is negligible), from Eq. \eqref{Pdot} we obtain 
$\vert \dot{M}/M \vert < 1.65 \times 10^{-11}/\text{yr}$. Then from Eq. \eqref{Mdot} 
we can derive an upper limit $\eps < 4.3 \times 10^{-13}$~eV, which nicely coincides with 
the result of Ref. \cite{Nussinov-new}. 
The pulsars J1141$-$6545 and J1952$-$2630 
discussed in~\cite{Nussinov-new} imply comparable bounds. 

Somewhat stronger limit can be obtained from the Hulse-Taylor  
pulsar B1913+16 known as a perfect binary system for testing General Relativity. 
The masses of the pulsar and its companion (presumably another NS)  
were determined with a great precision:  $M=1.4398(2)\,M_\odot$ and 
$M_c= 1.3886(2)\,M_\odot$, as well as the orbital period and its derivative: 
$P_b= 0.322997$~days and $\dot{P}_b = - 2.423(1) \times 10^{-12}$~\cite{Weisberg:2010}. 
For this compact binary, the predicted GW contribution is 
$\dot{P}_b^\text{GW} = -2.402531(14) \times 10^{-12}$  which almost saturates 
the observed value. After subtracting also the galactic correction 
$\dot{P}_b^\text{gal} = (-0.027\pm 0.005) \times 10^{-12}$, one obtains  
\be{Weisberg}
\frac{\dot{P}^{nn'}_b}{P_b} = \frac{\dot{P}_b - \dot{P}_b^\text{GW} - \dot{P}_b^\text{gal}}{P_b}
=(7.4 \pm 5.6) \times 10^{-13}/\text{yr}
\ee 
This, assuming  that both NS suffer mass loss, from \eqref{Pdot}
we get  $(\dot{M} + \dot{M}_c)/(M+M_c) 
= -(3.7 \pm 2.8) \times 10^{-13}/\text{yr}$ which can be interpreted as an upper limit 
\be{Weisb-eps}
\eps < 8 \times 10^{-14}~\text{eV}
\ee





Energy loss due to $n-n'$ transition can have interesting implications also for the NS 
surface temperatures. As we discussed in Sect. \ref{sec:evolution}, 
per every process $n n \to n n'$ in the NS the mirror neutron $n'$, 
due to the Pauli blocking,  
is produced with with typical energy $E_F = \xi^{2/3} \times 60$~MeV, 
and the energy production rate per baryon can be estimated as \eqref{Gamma-E}.  
According to Eq. \eqref{t-smallX}, for $\tau_\eps \gg t_U$ we have $X \ll 1$,  
so that all neutron stars can be considered as young,   
and  the energy production rate $\dot{E}_{nn'}$ in their interior is given by Eq. \eqref{Edot}. 
Since the M matter density is much smaller than the neutron density.  
the produced mirror neutrons will preferably decay as $n' \to p'e'\bar\nu'$  
producing a hot plasma of mirror protons and electrons gravitationally 
 trapped inside the neutron star,  though also various nucleosynthesis processes as 
should take place. 
Then, taking that the fraction $x$ of the produced energy is radiated 
via the thermal spectrum of mirror photons $\gamma'$,  
the ``mirror photosphere"  temperature $T_{\gamma'}$ of the NS can be estimated as 
\be{surface-Tpr}
T_{\gamma'} = \frac{\eps_{15}^{1/2} }{R_{12}^{1/2} } \left( \frac{\kappa M}{1.5\,M_\odot}\right)^{1/4} 
\! x^{1/4}  \times 3 \cdot 10^{5} ~\text{K} 
\ee
simply by equating $x \dot{E}_{nn'} = 4\pi R^2 \cdot \sigma T_{\gamma'}^4$, 
with $\sigma$ being the Stefan-Boltzmann constant.\footnote{The photon fraction $x$ 
is difficult to calculate precisely since the fraction $1-x$ is emitted in mirror neutrinos $\nu'$. 
For that one has  perform numerical simulations of mirror nuclear processes 
in the NS interior which depends on the  M matter density and thus on the age of the star.  
However, $T_{\gamma'}$ very mildly depends on $x$. 
 }

On the other hand, some part of energy produced in the NS will be emitted in terms of 
ordinary photons and neutrinos. Namely,  disappearance of the neutron in the reaction 
$nn \to n n'$ leaves the ``empty" level in the Fermi see which will be filled 
by transition of the neutron from the higher  level.  
Once again, since $nn\to nn'$ reactions take place close to the Fermi surface, 
this transition energy should be smaller than the Fermi energy $E_F$ by a factor of 50 or so 
as we obtained by the MC simulation.   
Thus, the ordinary component  should have less heating than the mirror one,  
and the NS surface temperature in terms of ordinary photons can be roughly estimated as   
\be{surface-T}
T_{\gamma} = \frac{\eps_{15}^{1/2} }{R_{12}^{1/2} } \left( \frac{\kappa M}{1.5\,M_\odot}\right)^{1/4} 
\!\times 10^{5} ~\text{K} 
\ee

The standard cooling mechanisms predict sharp drop of the temperature 
with the age of the star, leading to the surface temperature 
below $10^4$~K after $10^7$~yr, and below $10^3$~K after $10^8$~yr.
However, observations of some old pulsars detect that they are still warm, 
with the surface temperatures $10^5$~K or larger.  
For example, above discussed   PSR J0437$-$4715  
is the brightest millisecond pulsar in UV and X-rays. 
Its characteristic age is $\tau_c=3.2\times 10^9$~yr but its UV spectral shape 
suggests a thermal emission with the surface temperature 
$T_\gamma = (1.5\div 3.5) \times 10^5$~K \cite{Durant:2011}. 
Also PSR J2124$-$3358, a solitary 5\,ms pulsar of the age $\tau_c=3.8\times 10^9$~yr 
shows a thermal spectrum with $T_\gamma= (0.5\div 2) \times 10^5$~K \cite{Rangelov:2016}. 
On the other hand,  an younger pulsar B0950+08 ($\tau_c=1.8\times 10^7$~yr) 
also has a surface temperature $T_\gamma = (1\div 3) \times 10^5$~K \cite{Pavlov:2017}. 
These temperatures are much higher than predicted by cooling models, 
which means that some heating mechanisms operate in the NS. 
Namely, in the context of our model such surface temperatures can be 
explained by $n-n'$ transformation with $\eps \sim 10^{-15}$~eV. 
It should be noted, however, that the determination and interpretation of 
the NS surface temperatures are model dependent, namely they depend on interstellar extinction
and the models of pulsar magnetosphere, accretion from the partner and non-thermal emission, 
and there are other heating scenarios related to Urca processes with vortex friction and 
rotochemical reactions. Let us also note that observations of isolated slow pulsar J2144$-$3933 
($P=8.5$~s) with $\tau_c=3.3\times 10^8$~yr 
imply solely an upper bound $T_\gamma < 4.2 \times 10^4$~K  \cite{Guillot:2019} 
which disfavors some of heating mechanisms but still remains compatible with 
$n-n'$ transformation with $\eps < 2\times 10^{-16}$~eV or so. 
On the other hand, the suppressed thermal spectrum in PSR J2144$-$3933 
can be related also to environmental factors, and it would be premature to derive 
any serious conclusion  just on the basis of one non-detection. 

Finally, let us remark that for $\tau_\eps \gg t_U$ the mirror neutron stars can be 
in fact ``visible" in the UV diapason, with the surface temperature given by Eq. \eqref{surface-Tpr}. 
The energy produced in the mirror NS due to $n'-n$ transformation 
will be emitted in ordinary far UV photons with the rate \eqref{Edot}. For e.g.  
$\eps=10^{-14}$~eV this rate will be $\sim 10^{33}$~erg/s which is equivalent to the solar luminosity.

\subsection{Effects of fast NS to MS transformation}

Let us discuss now the situation when the NS transformation time is rather small, 
say $\tau_\eps < 10^{5}$~yr or so. In view of Eq. \eqref{tau-eps}, this would 
correspond to $n-n'$ mixing mass $\eps > 10^{-10}$\,eV or so.
Then the stars with the age larger than $\tau_\eps$ should be already transformed in the MMS, 
with (almost) equal amounts of ordinary and mirror baryons inside.   
Correspondingly, these stars will have no more energy losses due to $n'$ production 
and would not show the evolution effects discussed in previous subsection. 
In addition, also the compact objects initially born as mirror NS can be visible for us in this mixed 
form, and in addition they can be detected as ordinary pulsars provided that by 
some mechanism they acquire also the ordinary magnetic field. 
This can naturally be naturally induced by kinetic mixing term between ordinary and mirror 
photons $(\epsilon/2)F^{\mu\nu} F'_{\mu\nu}$ \cite{Holdom} which effectively makes M 
protons and electrons mini-charged (with ordinary electric charges $\sim \epsilon$). 
The cosmological bounds imply $\epsilon < 5\times 10^{-9}$ or so \cite{Lepidi} 
while the experimental limit from the positronium decays 
is yet weaker: $\epsilon < 5\times 10^{-8}$ \cite{Vigo}. 
As we noted above, the neutrons produced via $n'-n$ transition 
at first stages of the mirror NS evolution decay and produce electrons and protons 
which then enter in nuclear reactions with the neutrons forming nuclei.  
In any case, the mirror NS rotation, via Rutherford-like scatterings due to the photon kinetic mixing, 
 will drag the electrons rather than protons and ions, 
 inducing circular electric currents which can give rise to substantially large magnetic 
fields by the mechanism suggested in Ref. \cite{BDT}. In this way, it can become a 
complicated task how to distinguish between the old pulsars initially originated from ordinary and 
mirror NS since they should also have comparable surface temperatures. 
One possibility can be that ``mirror-born" pulsars should dominantly accrete mirror matter, 
since their companions in binaries should be M stars,  and thus they should be dominantly 
active in terms of mirror X-rays rather than in ordinary X-rays. 

A clear  phenomenological implication of the gravitational mass scaling \eqref{max-M} 
is that  for any EoS,  the last stable MMS mass  
is $\sqrt2$ times smaller then the last stable  mass of the ordinary NS.
Applying this relation 
to our first example (SLy EoS), $M_{\rm NS}^{\rm max} \approx 2.05\,M_\odot$  
implies $M_{\rm MMS}^{\rm max} \approx 1.45\,M_\odot$,  
while for the second (piecewise polytropic) example we have respectively
$M_{\rm NS}^{\rm max} \approx 2.57\,M_\odot$ and 
$M_{\rm MMS}^{\rm max} \approx 1.81\,M_\odot$. 
Thus, the existence of heavy pulsars with 
$M\simeq 2M_\odot$ and characteristic ages $ \tau_c > 10^9$~yr  
demand rather stiff EoS allowing $M^{\rm max}_\text{NS} \simeq 2.7~M_\odot$ or so. 

As already discussed, the scaling relations do not hold for stars 
with the same total baryonic number.  
Therefore the configurations related by the $\sqrt{2}$--scaling  of Eq.~\eqref{sqrt2} 
and preserving the stellar compactness do not correspond to different evolution stages of the same star. 
A young neutron star with a mass $M_\text{NS}$ exclusively composed of O component 
with central density $\rho_\text{cO}$, 
due to $n\to n'$ conversion  slowly evolves to a more compact MS, with  
$\rho_{cM}(t)\neq 0$ and $\chi(t) = \rho_{cM}(t)/\rho_{cO}(t)$ adiabatically increasing in time.
 If this way,  the star continuously converts gravitational energy in heat, 
which should be emitted presumably in terms of ordinary and mirror neutrinos and photons. 
As depicted in Fig.~\ref{fig:Sly}, during the conversion the star becomes more compact, 
and its mass becomes somewhat smaller due to the gain in the gravitational binding energy. 
The mass difference between the initial and final states is about 
$0.1~M_\odot$ for both  considered EoSs.  This means, that e.g. for the case of the SLy EoS, 
a NS with the initial mass $M_\text{NS} = 1.5~M_\odot$ and  radius of about $12$ km, 
can evolve into an asymptotic  final configuration of a MMS ($\chi=1$) with 
a slightly smaller mass ($M \simeq 1.4~M_\odot$)  but considerably more compact, 
with the radius of about $8$ km or so.   
Therefore, if the observations will find two compact stars 
with  masses of e.g. $1.4~M_\odot$ but having very different radii, 
such objects can be interpreted as two MS  with different ``mixtures'' of the O and M components 
(i.e. with  different values of $\chi$) and thus with different ages.  
Unfortunately, for the compact stars whose masses are known with a high   
precision the radii remain practically unknown, and few cases when both masses and radii 
can be both determined have very poor precision \cite{Ozel}.
 However, if the precise radius measurements  by the  NICER mission, 
 see for instance~\cite{Ozel:2015ykl},  yield  very different radii 
  irrespectively of the objects mass,  one could interpret this result as two  MSs 
  in different stages of the evolution: 
  the more compact star should be older than the less compact one.    

Quite interestingly, recent analysis of NS masses in binary systems suggest a bimodal distribution~\cite{Schwab}. 
 A possible interpretation is that the two distributions correspond to different type 
of stars,  with one (lighter) component corresponding to standard NS and another (heavier) 
component  corresponding to twin configurations made of hybrid stars (HS)  
with a deconfined quark matter core or entirely quark stars (QS).  
However, the quark stars can only cover a restricted region of the mass-radius diagram.  
In our picture, instead, the MS evolution allows to span a larger region of the  mass-radius diagram. 
  
 Remarkably, the $n \to n'$  conversion process could affect the total  distribution 
 of the NS masses.  For definiteness, let us focus on the SLy EoS.  
First, we note that NS born with mass $M_{\rm NS} < 1.55\,M_\odot$ 
continuously decreases its gravitational mass and become the MMS after a time $t> \tau_\eps$.   
Second, NS with larger initial masses, $M_{\rm NS} > 1.6~M_\odot$, cannot become MMS. 
In this case the star is doomed to collapse to a black hole. 
A neutron star with initial mass $M = 1.8~M_\odot$ 
will become unstable and collapse to a BH in a time $t\simeq 0.1\,\tau_\eps$ after the birth,  
while the NS born very close to the last stable configuration would collapse much faster. 
Both effects should alter  the NS mass distribution predicted by the supernova explosion 
mechanism, reducing the number of compact stars with large masses while increasing 
the number of stars with low masses, as was discussed in~\cite{Massimo}.
Unfortunately, the situation is more complex because one cannot exclude the possibility 
that due to the matter fall-back after supernova explosion the NS can transformed 
into the HS or QS \cite{BBDFL} (see also discussion in \cite{Drago:2020}).  
In general, standard evolution processes may play an important role, 
leading to a $(0.2 - 0.3)\, M_\odot$ increase of masses due to matter fallback 
after the supernova explosion or accretion in recycled  NS~\cite{Ozel:2012}. 
Which final state, HS or QS, will be reached  in this transition depends 
on the details of the quark matter EoS and on the amount of accreted mass. 
All these mechanisms lead to the mass increase,  
thus it is generally believed that  the mass of  NS after their birth can only increase, 
while the ordinary to mirror matter conversion should reduce 
the gravitational mass.\footnote{On the other hand, 
 also the NS with estimated masses $\simeq 1 M_\odot$ have been observed~\cite{Ozel:2012}, 
which challenges the present understanding of core-collapse 
neutron star formation~\cite{Lattimer}.} 
 
The NSs in double neutron star (DNS) binaries represent an interesting case for studying 
the effect of the $n \to n'$ conversion on the mass distribution of compact stars. 
These NSs  are thought to have received little or no accretion, 
thereby  reflecting the NS stellar mass at birth with a direct  link to the supernova mechanism. 
The observed DNS masses  can be nicely fitted with a gaussian with a central value at 
$\approx 1.4\,M_\odot$ and rather small dispersion, $\sigma \approx 0.05\, M_\odot$~\cite{Ozel}, 
 though it is not quite clear why a general supernova explosion 
 should lead to such a central mass and to such a peaked mass distribution. 
 Indeed, the gravitational mass for collapsed cores could be less than $1.3\, M_\odot$, 
 but since in a DNS accretion can only happen by matter fallback, a large increase 
 in the dispersion of NS masses is expected~\cite{Zhang}.
   
 The effect of  $n \to n'$ conversion process is to shift the peak of the mass distribution 
 at birth towards lower values~\cite{Massimo}. 
 Therefore, if one could prove that the NS mass distribution at birth is wider and extends  
 to values  significantly above than $1.5\, M_\odot$, 
 then the narrow distribution of masses in the DNS systems 
 could evidence of the effect of the $n \to n'$ conversion process. 
 Once again, in the context of the Sly EoS, this distribution in fact 
 should be non-gaussian with a cutoff at maximum mass $M^\text{max} \simeq 1.45~M_\odot$ 
 for older stars, with ages exceeding $\tau_\eps$. 

As for younger stars, they still can be in the transformation process, and thus can have 
different radii. In addition, the evolution can manifest in observational phenomena  
as e.g. pulsar ``glitches", sudden increase of the rotation frequency caused by irregular  
transfer from the NS interior to the crust and by the after ``star quake" 
rearrangement of the crust.  In fact,  $n-n'$ conversion can proceed only in the NS interior 
which adiabatically shrinks the neutron liquid while this shrinking is not adiabatically 
followed by the shrinking of the rigid crust which then ruptures in discrete events. 
Depending on the situation, such effects can cause also pulsar ``anti-glicthes", 
events of a sudden spin-down.   It is tempting to think that also the phenomena 
of soft gamma repeaters or intermittent pulsars can be related to the effects of 
$n-n'$ conversion during the evolution.  


Finally, let us comment on the combined effect of conversion to mirror matter 
and formation of strange quark matter at  high  nuclear densities. 
According to the Bodmer-Witten hypothesis \cite{Bodmer} strange quark matter 
can be the  energetically favored ground state at large densities. 
Thus, the NS whose masses can strongly increase by accretion  reaching a critical value, 
say $M_\text{th}\simeq 1.6\,M_\odot$,  
can be promptly transformed  into a star made at least in part of deconfined quark matter \cite{BBDFL}.  
Which final state, HS or QS, will be reached  in this transition depends on the amount of accreted mass
as well as on the details of the quark matter EoS. 
In quark matter the  transformation to mirror matter should be suppressed for two reasons, 
first because there are not much neutrons to transform, and second because 
quark matter is self-bound (as standard nuclei), and therefore  
the transition to mirror nuclear matter should give no energy gain, 
in particular if quark matter is in a color superconducting phase~\cite{Alford:2007xm, Anglani:2013gfu}. 
Therefore, in QSs almost entirely consisting of quark matter will not be transformed into mixed stars. 
The case of HS is more interesting since $n-n'$ transition will be still effective in the part of star 
consisting of the neutron liquid. Therefore, in a time $\tau_\eps$ it will be transformed into a  
QS with a core of M neutrons in fact forming a mirror NS inside the QS. 
For rather heavy stars, in which also the density of the mirror core can reach the threshold value, 
the mirror neutrons core can in turn transform into quark matter, thus forming a mixed quark star. 

Therefore, also the strange QS can have some M cores  
consisting of mirror neutrons or deconfined mirror quarks, depending in their mass and 
evolution history. 
Reciprocally mirror QS can have ordinary matter cores 
which can be detectable by their electromagnetic radiation. 
Once again, mirror stars with small enough initial masses 
should consist entirely of M neutrons and thus they should evolve into the MMS.

\section{Neutron star mergers and associated signals} 
\label{sec:GW}

Let us briefly discuss the implications on $n-n'$ transitions for gravitational wave (GW) 
bursts from the NS mergers and the associated electromagnetic signals as 
gamma ray burst (GRB) and kilonova events which are also known 
as the main source of production of heavy (trans-iron) elements 
in the Universe.  

LIGO Collaboration detected two candidates. 
The first event GW170817 \cite{GW170817} is considered as a clear signal of 
the ordinary NS merger, with masses of two stars $M_{1,2}$  compatible with $\simeq 1.4\,M_\odot$ 
and their total mass $M_\text{tot} = (2.75 \pm 0.02)\,M_\odot$  typical for NS binaries.  
Remarkably, the GW signal was accompanied also by a weak GRB  
as well as by electromagnetic afterglows in different diapasons. 

The second candidate GW190425 \cite{GW190425} is more unusual. 
While the best fit masses of individual components 
($M_1 \simeq 1.8\,M_\odot$ and $M_2 \simeq 1.6\,M_\odot$)   
are within the mass range of the observed NS, 
both the source-frame chirp mass $(1.44 \pm 0.02)\,M_\odot$ 
and the total mass $(3.4 \pm 0.2)\,M_\odot$  of this system are significantly larger 
than those of any other known NS binaries. 
In addition, no confirmed electromagnetic event has been identified in association 
with this GW signal which suggests that this event could be originated by the merger 
of mirror NS \cite{Addazi-Rezi}. 

 The possibility of $n-n'$ conversion adds new features to the picture. 
Namely, if the conversion time $\tau_\eps$ is short, then the old neutron stars 
should exist today in the MMS form, with equal amounts of the O and M components inside. 
Therefore, the MMS mergers  should have potentially observable electromagnetic  counterparts 
irrespectively of their origin (ordinary or mirror). 
In fact, one cannot exclude the possibility that GW170817 was induced by coalescence of stars 
which were initially born as the mirror NS and then evolved into the MMS. 

As for GW190425, 
the location of this merger is practically unknown since the GW
was essentially detected only by the LIGO Livingston interferometer: the LIGO Hanford 
 at this moment was off-line while the signal of Virgo was at the level of noise. 
Thus, the non-detection of the associated electromagnetic counterpart  
does not really exclude that it was a merger of the two MMS. 
In addition, one can also speculate about the unusual mass parameters  of this system 
assuming that it was born as a mirror NS binary. 
As far as mirror world should be helium dominated  \cite{BCV} and so the 
evolution of M stars and their pre-collapse conditions can be different from 
those of ordinary matter stars \cite{BCCP}, then core-collapse of mirror progenitors 
could produce the NS with somewhat larger birth masses. One can also 
hypothesise that GW190425 was a coalescence of M quark stars 
with null or small amount of ordinary matter inside, and so without potentially
detectable electromagnetic association. 

 However, the enhanced compactness of the MMS with respect to NS 
can have interesting implications.  
Namely, the GW signal from the NS coalescence is sensitive to  
the tidal deformations that each components gravitational field induces on its companion,  
and thus it can give relevant information about the EoS describing the NS and their radii. 
In particular, the analysis of  GW170817 waveform \cite{Abbott:2018} favors the softer EoS as Sly \cite{Sly} 
rather than the stiffer ones as e.g. \cite{Mueller}. 
For the Sly EoS, assuming that the component masses are $1.4\,M_\odot$, 
the limits on tidal deformability obtained from GW170817 waveform analysis 
 implies the limit on their radii $R_{1,2} > 10$~km or so \cite{Abbott:2018}. 
 This is somewhat larger than
  the MMS radii $R_\text{MMS} = 8.5$~km predicted by the Sly in view of scaling relation 
\eqref{R-scaling}. However, it should be premature to make strong conclusions from 
this discrepancy before solid statistics is achieved on the GW signals,  
moreover that other interpretations were also discussed which  
allow rather small radii \cite{Burgio:2018}.  


If the transition time is larger than the age of the universe, $\tau_\eps \gg t_U$, 
then only a  small fraction of ordinary nuclear matter can be transformed into 
 M matter in the NS.  
 The produced mirror matter should form a small core inside the original star.  
This core should only be  bound  by gravity with no material friction with the dominant ordinary 
component. Therefore, every sudden collision with external bodies or fast accretion 
of a large chunk of matter could cause relative vibrations between the two components, 
which may manifest as some sort of glitches. In any case, this should contribute 
to the deformability of the compact star, which can be tested by analyzing the GW waveforms 
from the NS mergers. Further, when the dominant O components of a merging binary system   
hit each other and coalesce, the smaller mirror cores should continue 
their rotation for some time and merge with some delay, eventually  producing some 
non-trivial  perturbations in the GW waveform. However, if the mirror cores have masses 
smaller than the evaporation limit ($0.1 M_\odot$ or so) and 
moreover if their orbiting ellipse has a high eccentricity,  they can be thrown away from the 
merger site by a sling-like effect and then explode due to decompression. 
This can giverise to the kilonova-type  events  producing a hot mirror neutron rich 
cloud which can be at the origin of r-processes  producing heavy elements in the
mirror sector.  

A different and intriguing possibility is related  to the mergers of mirror neutron stars.  
 in the absence of $n-n'$ mixing, the mirror NS will suffer no conversions into ordinary 
 matter and thus  the mirror NS mergers would produce  gravitational wave signals  
not accompanied by any standard electromagnetic counterpart \cite{Addazi-Rezi}. 
 In this case the mirror NS mergers will look as `invisible but not silent'.  
 But If the mirror star hosts an ordinary nuclear matter, 
 meaning that it is a MS with dominant in mass (and larger in size) M component 
 having a small core of  standard nuclear matter,
 then after the merger of dominant  M components  the  subdominant  ordinary cores 
 could continue the orbiting for a while and then, via their collision or decompression,  
 give rise  to a hot neutron-rich cloud around the coalescence site.  
In other words, we suggest that the observed  kilonova events as well as  gamma ray bursts,
 or at least some of them,  
could originate from the merging of MSs with a dominant mirror component.  
The existence of MSs with a dominant M component may have a number of 
additional phenomenological effects, in particular if their ordinary cores produced 
by $n'-n$ transitions  develop substantial magnetic fields,  they could be observed  as ordinary pulsars. 

Another  intriguing possibility is that  mirror matter has the baryon asymmetry 
of the opposite sign to ordinary matter,  so that mirror neutrons inside transform 
into the standard anti-neutrons rather than into the neutrons. 
Such a situation can be naturally realized in co-baryogenesis models between O and M 
sectors discussed in Refs. \cite{ADM-IJMP,ADM}. In this case a mirror anti-neutron star
would  develop in its interior a core consisting of ordinary antimatter. The gravitational 
merging of such mirror star binaries can be at the origin of anti-r processes 
 which would produce ordinary anti-nuclei, and anti gold in particular.  
 Electromagnetic signals of such anti-kilonovae cannot be distinguished from ordinary kilonova events,
 but the produced anti-nuclei can be hunted by the AMS2 Collaboration in the spectrum of cosmic rays.  
In addition, if reasonably large magnetic fields can be transferred to the 
rotating ordinary anti-core, such a star  can be seen as a pulsar. 

\section{Conclusions}
\label{sec:conclusions}

We have discussed the possibility that the ordinary neutron stars, 
via $n-n'$ conversion,  can develop the mirror matter cores in their interior. 
These cores gradually increase in time and only the stars with masses than 
some (the EoS dependent) critical value  can survive asymptotically in time reaching 
the maximally mixed configuration while the heavier ones should collapse into black holes. 

To distinguish from other works, let us remark that possibility of the 
neutron stars with small mirror cores formed by the dark matter accretion 
was discussed in Refs.~\cite{Sandin}. However, dark matter 
accretion rate cannot be very effective and it can destabilize only heaviest pulsars 
with masses already very close to maximum mass allowed by the given EoS. 
In fact, the neutron star with a bigger rate would accrete the normal matter 
in which interior it was born, as in is the case for recycled neutron stars. 

The neutron star conversion into dark neutron star via the transitional magnetic moment 
induced by the neutron--dark neutron mixing was discussed in Refs.~\cite{McKeen}. 
The dark neutron was considered as an elementary particle with mass closely 
degenerate to the neutron mass, within 1 MeV or so, and without significant self-interaction.  
It was shown that such dark stars cannot have masses larger than $0.7~M_\odot$. 
For stabilizing these objects, the dark neutron self-interactions were {\it ad hoc} 
 introduced in Ref.~\cite{Cline}, which possibility for compact dark matter stars 
 was previously  studied in Ref.~\cite{Narain}. 

In our case,  once the concept of mirror matter is adopted, 
there is no need for complementary hypotheses for maintaining the stability of 
the mixed stars, since the mirror nuclear matter should have exactly the same EoS as the ordinary one. 
Therefore, the existence of the maximally mixed stars, with evenly distributed O and M components, 
implies only the upper limit on their maximum mass \eqref{max-M}. 
However, this limit also depends on the chosen EoS (and it can also  be avoided 
by assuming that the heavy pulsars are in fact the quark stars, since $n-n'$ transition 
should be ineffective in quark matter). 

We have discussed various astrophysical implications of this scenario. 
First we discussed the situation when the NS transformation time into the MS is 
larger than the universe age, $\tau_\eps> t_U$ in which case all existing NS 
should still be in the evolution processes. We have shown that no astrophysical limits 
related to the pulsar characteristic ages, orbital period change in binary pulsars or 
the pulsar surface temperatures, exclude the possibility of $n-n'$ mixing \eqref{n-npr}
with $\eps < 10^{-15}$~eV (corresponding to oscillation times
$\tau_{nn'} = \eps^{-1} > 1$~s or so) which is the target of several ongoing and planned 
experiments for searching $n-n'$ oscillations via the neutron disappearance ($n\to n'$) 
or regeneration ($n\to n'\to n$)  in the minimal picture of mirror world with exact $Z_2$ parity. 
In this case $n-n'$ oscillation can have also interesting implications for the 
ultra-high energy cosmic rays. 
 
 Neither the possibility of $\eps > 10^{-10}$ can be excluded. Also this parameter area   
 has  phenomenological interest since $n-n'$ oscillation could solve the neutron lifetime problem 
 provided that O and M neutrons have a mass splitting of few hundred neV \cite{lifetime},  
 and experiments are underway for testing it. 
 It implies the NS transformation times $\tau_\eps < 10^5$~yr or so 
 in which case the observed NS with typical ages $10^6\div 10^{10}$~yr 
 should be already transformed to the MMS configuration. 
 In this case no evolution effects can be manifested by old pulsars while 
 for pulsars younger than $10^5$~yr are still hot, 
 with intrinsic temperatures larger than $10^7$~K and thus compatible with the limits 
on heating produced by $n-n'$ conversion.  

The intermediate range $\eps = (10^{-15} \div 10^{-10}) \,\text{eV}$ 
is more subtle. Namely, for $\eps \leq 10^{-13}$~eV we have $\tau_\eps \geq t_U$ 
the interval $\eps = (10^{-14} \div 10^{-13}) \,\text{eV}$ is disfavored by the limits 
on the pulsar and pulsar binary timings and on the NS surface temperatures. 
As for the interval  $\eps = (10^{-13} \div 10^{-10}) \,\text{eV}$, 
 the effects of $n-n'$ transition will be dependent on the NS age and 
 a careful analysis is needed to determine the upper edge of the excluded area.  

We have also briefly discussed the effects of $n-n'$ conversion in hybrid quark stars 
and for the gravitational mergers in binary systems.   Interesting possibility is that 
also the coalescence of mirror-born neutron stars could give rise to 
the weak GRB and associated kilonova events.  

\bigskip 

{\bf Acknowledgements} 
\medskip

\noindent
The work of Z.B. was supported in part by the research grant 
``The Dark Universe: A Synergic Multimessenger Approach" No. 2017X7X85K 
under the program PRIN 2017 funded by the Ministero dell'Istruzione, Universit\`a e della Ricerca (MIUR),  
and  in part by Shota Rustaveli National Science Foundation 
(SRNSF) of Georgia, grant DI-18-335/New Theoretical Models for Dark Matter Exploration. 
Part of this work was previously reported by M.M. at the Nordita Workshop ``Particle Physics with Neutrons at the ESS", 
Stockholm, Sweden,  10-14 Dec. 2018, 
and by Z.B. at Spontaneous Workshop XIII ``Hot Topics in Modern Cosmology", Cargese, France, 
5-11 May 2019. 

 \end{document}